\documentclass[aps,prx,twocolumn,floats,showpacs,superscriptaddress,nofootinbib]{revtex4-1}
\usepackage{graphicx,epsfig}
\usepackage{times,bbm}
\usepackage{graphics,dcolumn,bm,float}
\usepackage{amssymb,amsmath,rotate,color}
\usepackage[title,titletoc,toc]{appendix}
\usepackage{mathtools}
\usepackage{booktabs}
\usepackage{tcolorbox}

\usepackage[pagebackref=false,colorlinks,linkcolor=blue,citecolor=cyan,urlcolor=blue]{hyperref}

\begin{document}
	\unitlength 1 cm
	\newcommand{\be}{\begin{equation}}
		\newcommand{\ee}{\end{equation}}
	\newcommand{\nn}{\nonumber}
	\newcommand{\vk}{\vec k}
	\newcommand{\vp}{\vec p}
	\newcommand{\vq}{\vec q}
	\newcommand{\vkp}{\vec {k'}}
	\newcommand{\vpp}{\vec {p'}}
	\newcommand{\vqp}{\vec {q'}}
	\newcommand{\bk}{{\vec k}}
	\newcommand{\bp}{{\bf p}}
	\newcommand{\bq}{{\bf q}}
	\newcommand{\br}{{\bf r}}
	\newcommand{\bR}{{\bf R}}
	\newcommand{\bsq}{{\boldsymbol{q}}}
	\newcommand{\bsk}{{\boldsymbol{k}}}
	\newcommand{\bsS}{{\boldsymbol{S}}}
	\newcommand{\bss}{{\boldsymbol{s}}}
	\newcommand{\up}{\uparrow}
	\newcommand{\down}{\downarrow}
	\newcommand{\cdag}{c^{\dagger}}
	\newcommand{\hlt}[1]{\textcolor{red}{#1}}
	\newcommand{\ac}[1]{\textcolor{green}{#1}} 
	\newcommand{\ba}{\begin{align}}
		\newcommand{\ea}{\end{align}}
	\newcommand{\la}{\langle}
	\newcommand{\ra}{\rangle}
	\newcommand{\bmt}{\left[\begin{matrix}}
		\newcommand{\emt}{\end{matrix}\right]}
	\newcommand{\bearr}{\begin{eqnarray}}
		\newcommand{\eearr}{\end{eqnarray}}
	\newcommand{\eps}{\varepsilon}

	\title{Optical conductivity of triple point fermions}
	\author{Alireza Habibi} 
	\affiliation{Department of Physics, Sharif University of Technology, Tehran 11155-9161, Iran}
	\author{Tohid Farajollahpour} 
	\email{tohidfrjpr@gmail.com}
	\affiliation{Department of Physics, Sharif University of Technology, Tehran 11155-9161, Iran}
	\author{S. A. Jafari}
	\email{jafari@physics.sharif.edu}
	\affiliation{Department of Physics, Sharif University of Technology, Tehran 11155-9161, Iran}
	\affiliation{Center of excellence for Complex Systems and Condensed Matter (CSCM), Sharif University of Technology, Tehran 1458889694, Iran}
	
	\begin{abstract}     
		As a low-energy effective theory on non-symmorphic lattices, 
		we consider a generic triple point fermion Hamiltonian which is parameterized by an angular parameter $\lambda$. 
		We find strong $\lambda$ dependence in both Drude and interband optical absorption of these systems.
		The deviation of the $T^2$ coefficient of the Drude weight from Dirac/Weyl fermions can be used as a quick way to optically distinguish the triple point degeneracies from the Dirac/Weyl degeneracies. 
		At the particular $\lambda=\pi/6$ point, we find that the "helicity" reversal optical transition matrix element
		is identically zero. But deviating from this point, the helicity reversal emerges as an absorption channel. 
	\end{abstract} 
	\pacs{
		78.20.−e,  
		73.43.Cd, 
		71.90.+q, 
	} 
	\maketitle
	
	\section{introduction}
	Linear band crossings in solids usually correspond to Dirac/Weyl equation~\cite{Armitage,Wehling}. 
	The continuum limit of the effective theory of electrons in these systems becomes the Dirac/Weyl equation which is analogous to the Dirac/Weyl equation in high energy physics. However, for solids with non-symmorphic
	crystals, it may be possible to find forms of electronic theories which have no counterparts in high energy physics~\cite{Bradlyn}. 
	In fact, the appearance of strange forms of fermionic quasiparticles disobeying spin-statistics theorem is not the only consequences of such systems. Indeed for those nonsymmorphic lattices supporting tilted Dirac cone, a covariant
	description can be achieved only with a metric which deviates from the Minkowski metric~\cite{jafari2019electric,farajollahpour20198}.
	
	If the symmetries of spacetime were restricted to Pioncar\'e, the fermions would be obliged to have half-integer spins. However, on non-symmorphic lattices, the Lorentz (and hence Poincar\'e) symmetry is not obeyed by the effective theory. As such, there can be fermionic excitations that can be described by a deformation of $S=1$ theory. These are called triple fermions~\cite{Bradlyn} which are somehow intermediate between two-fold and four-fold degeneracies~\cite{DaiCoexixt,DaiTriply}. As such, they have no counterpart in high energy physics, simply because there are no three-dimensional representations of the Clifford algebra. 
	In addition to three-fold degeneracies, six and eight-fold fermions degeneracies can also arise
	on nonsymmorphic lattices~\cite{Soluyanov}. 
	Recently Fulga and Stern have pointed out that the triple fermions are not limited to the non-symmorphic models and minimal models with symmorphic symmetries that host triplet fermions can be constructed~\cite{stern}.  
	Hu and coworkers have suggested that the interplay between spin-tensor and spin-vector-momentum coupling can give rise to three types of triple point fermions classified by different monopole chargers~\cite{spin}. 
	
	The focus of the present paper is to understand the optical conductivity of materials that support triple femions~\cite{Bradlyn}.
	In this family of materials which are based on the one nodal line (type A) or four nodal lines (type B), 
	triple fermions have two distinct topological natures~\cite{Soluyanov}. Theoretical predictions 
	propose $\mathrm{ZrTe}$ family of materials as type A triple points topological semimetals~\cite{Soluyanov}, whereas 
	type B triple points are realized in $\mathrm{CuPt}$-ordered $\mathrm{InAs_{0.5}Sb_{0.5}}$~\cite{typeBexp1}, 
	and also in $\mathrm{HgTe}$ samples which are strained along $(1~1~1)$ direction~\cite{typeBexp2}. 
	The triple fermions belonging to crystallography group no. 199~\footnote{In this paper we consider the effective theory of group no. $199$.} 
	are classified by Chern number. 
	A candidate for the group 199 is $\mathrm{Pd_3Bi_2S_2}$ that exists in single crystal form~\cite{WEIHRICH}. 
	Among the interesting properties of this family of triplet point fermions, 
	they support Fermi arcs~\cite{Soluyanov}. Therefore the Fermi arcs are not limited to Weyl materials~\cite{Weylexp1,Weylexp2,Weylexp3,Faraei2018Green,Faraei2019}.
	According to {\em ab-initio} calculations triple point fermions are predicted to exist in several half-Heusler compounds~\cite{Heusler} and furthermore, they support Fermi arcs. 
	Other interesting physical phenomena associated with triple point fermions include topological Lifshitz transitions~\cite{Soluyanov} and anomalies in transport~\cite{anomalus}.

	\begin{figure*}[!tbh]
		\centering
		
		\includegraphics[width=0.3\linewidth]{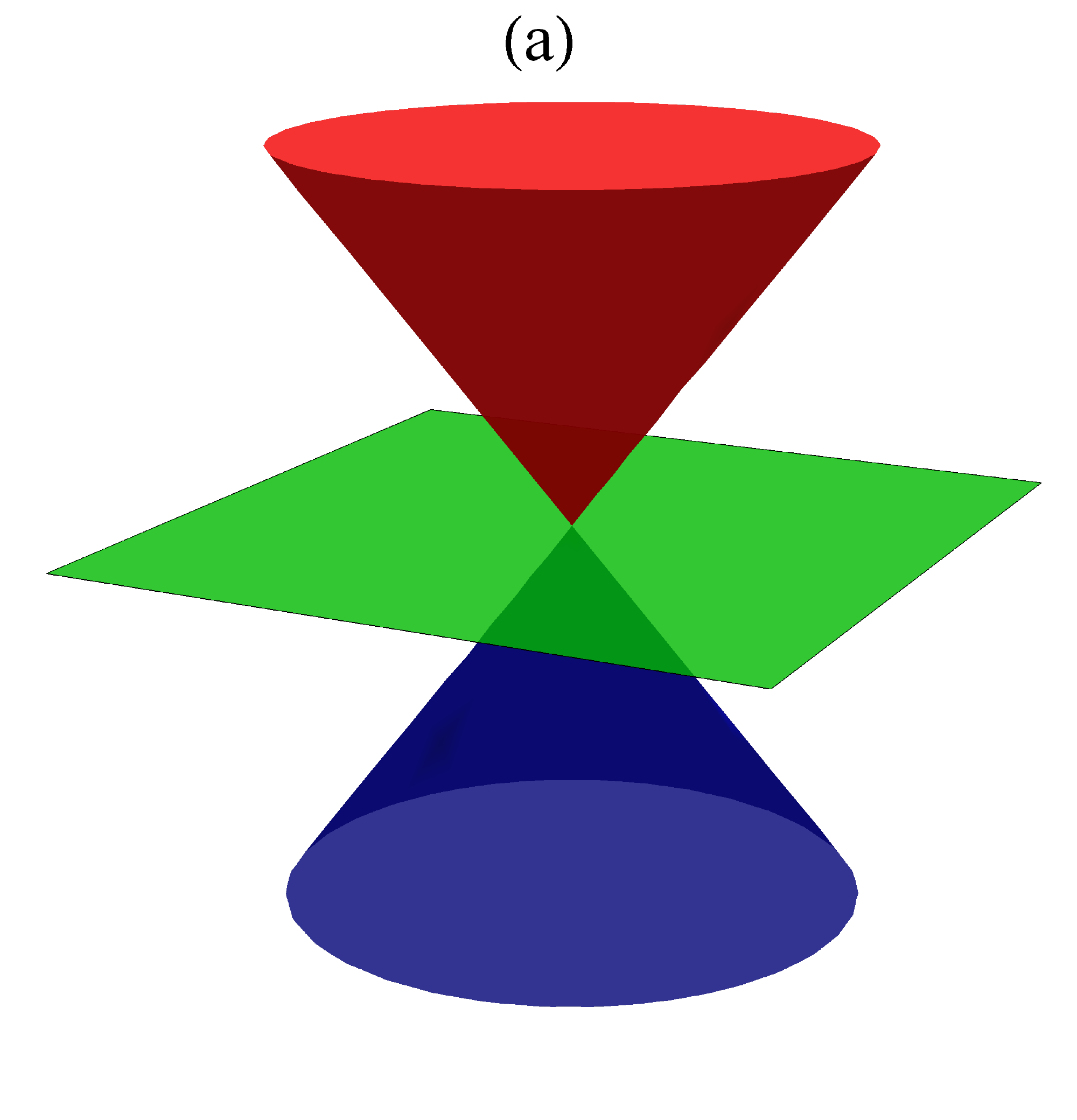}
		\includegraphics[width=0.3\linewidth]{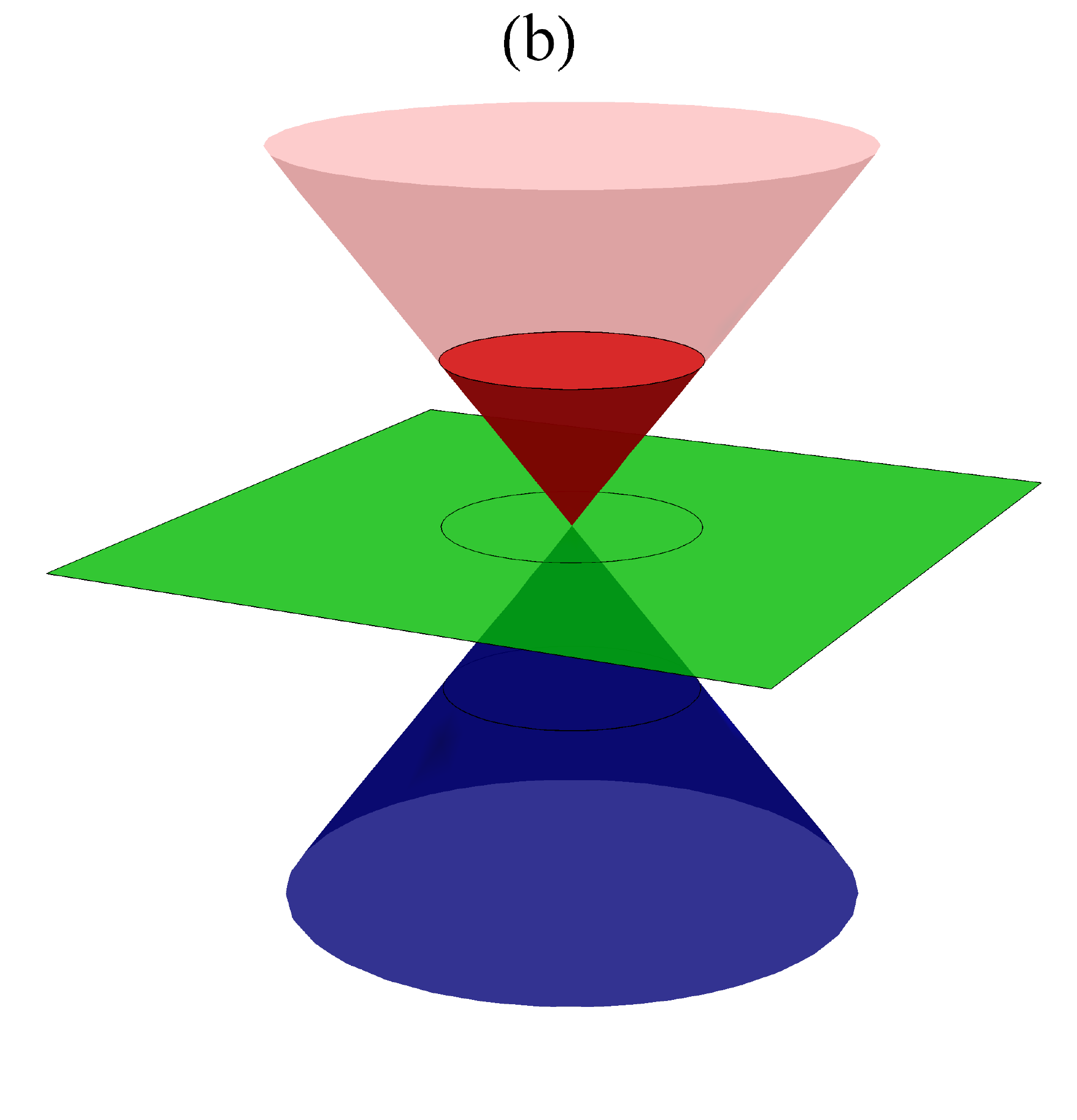}
		\includegraphics[width=0.3\linewidth]{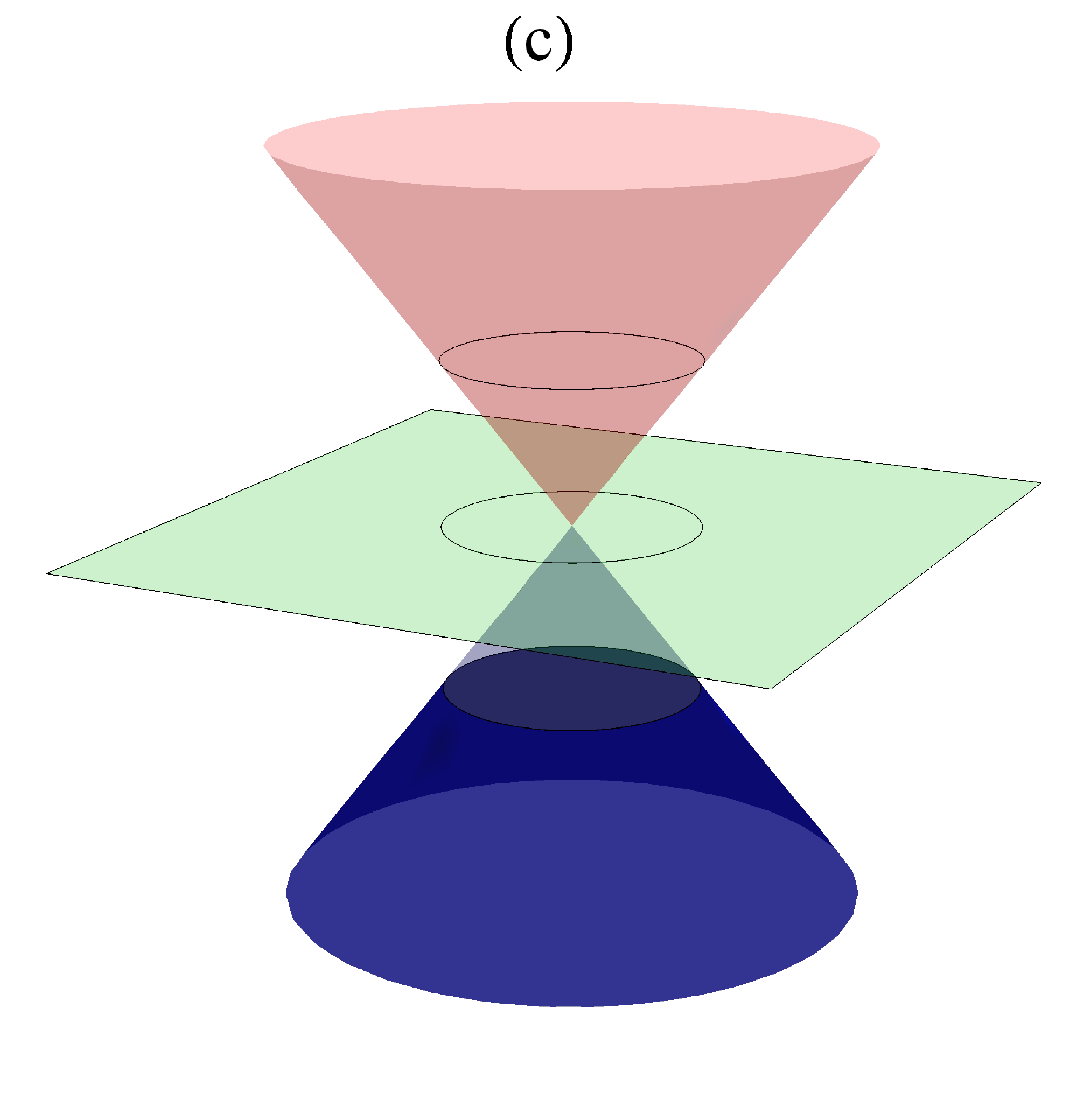}
		\caption{(Color online)  Schematic representation of triple fermions dispersion relation for $\lambda=\pi/6$. (a) The flat green band and Dirac cone altogether make a triple band model. Tunning of chemical potential has been illustrated in (b) and (c).  The gray circle is the projection of chemical potential at $k_z=0$ plane.  }
		\label{fig:ek}
	\end{figure*}
	
	In this paper, by means of Kubo formula, we  
	obtain various component of the optical conductivity tensors $\sigma_{\alpha \beta}$ 
	for the three band model, Eq.~\eqref{kdotS.eqn}. This equation is parameterized by 
	a phase variable $\lambda$ that defines $a=e^{i\lambda}$ in Eq.~\eqref{kdotS.eqn}. 
	At $\lambda=\pi/6$ simple closed-form expressions for the components of the optical conductivity tensor can be obtained. Away from this point, numerical computations are straightforward. For this model, it turns out that the $xy$, $xz$ and $yz$ components of optical conductivity tensor vanish which implies that there is no Hall conductivity in the sample. 
	
	The energy spectrum of triple fermions is composed of positive and negative energy bands which are similar to Weyl semimetals.
	However, there is an additional flat band in the middle which has no counterpart in Dirac/Weyl materials. So it is instructive to
	compare the optical absorption of Weyl and triplet point fermions. 
	The dependence of optical response to temperature and frequency in Weyl semimetals (TaAs) has been studied in Ref. \cite{Xu}. Their study reveals a narrow Drude response with a linear frequency dependence. They also found that by cooling, the weight of Drude peak decreases as $T^2$. 
	Even signatures of the chiral anomaly of Weyl semimetals can be traced in their optical absorption spectrum~\cite{carbotte}. 
	So everywhere in this paper, we contrast our results against the optical lineshape of Weyl semimetals emphasize the features that are peculiar to triple fermions. It should be noted that the optical and Hall conductivity of three band models in two-dimensions are studied in the context of the dice lattices~\cite{dice4,dice1,dice2,dice3}, while our system is three dimensional. 
	
	The paper is organized as follows: In the Sec.~\ref{Hamiltonian} we consider a $\bsk . \bsS$ Hamiltonian for which we construct the Green function. 
	From these Green's functions, the optical conductivity tensor is calculated in sec.~\ref{op1}. 
	We end the paper by summary, discussions and concluding remarks in Sec.~\ref{results}.
	
	\section{Hamiltonian and method}\label{Hamiltonian}
	The generic model of band touching is $\bsk.\bsS$ where $\bsS$ are angular momentum matrices.
	In the case of Dirac/Weyl materials, $\bsS$ are Pauli matrices, or their direct products~\cite{ZeeQFTBook}. 
	In the case of non-symmorphic lattices, however, $\bsS$ can be spin-$1$ or spin-$\frac{3}{2}$ matrices depending on the dimension of irreducible representations of its space group. For example, the space group number $199$ that harbors three-dimensional representation at $P$ point of the Brillouin zone, the angular momentum matrix $\bsS$ will correspond to spin-$1$. 
	Details of this group (no. 199) and derivation of the effective Hamiltonian is discussed in Ref.~\cite{Bradlyn}. 
	It should be mentioned that, existence of three-fold (six-fold or eight-fold) degeneracies has long
	been known from a band theory perspective~\cite{bradleyBook,kovalevBook}. 
	Therefore the most general $\bsk . \bsS$  Hamiltonian for space group number $199$ which is allowed by symmetry to first order 
	in $\bsk$ can be written as, 
	\begin{align}
		H_{199}=\begin{pmatrix}
			0&  a k_x&a^* k_y  \\ 
			a^* k_x& 0 &  a k_z\\ 
			a k_y& a^* k_z & 0
		\end{pmatrix} ,
		\label{kdotS.eqn}
	\end{align}
	where $a=|a|e^{i\lambda}$ and $a$ has dimension of $\hbar v_F$ with $v_F$ being appropriate Fermi velocity. 
	In this paper we set $\sigma_0=e^2/\hbar$ and $\hbar= v_F=1$. Although at any $\lambda$ one can obtain closed form expressions for the
	eigenvalues and eigenstates of this Hamiltonian (see Appendix~\ref{eigen.sec}), at $\lambda=\pi/6$ (or any other
	$\lambda$ that satisfies $\cos3\lambda=0$), these expressions becomes particularly simple. So most of the analytic
	results will be presented for $\lambda=\pi/6$. We will then compare the optical absorption at this particular point
	with other $\lambda$ values. Away from $\bsk=0$ the bands are 
	non-degenerate and the exception is $\lambda=\frac{n\pi}{3}$ ($n$ is integer) where 
	the bands are degenerate along the lines $|k_x|=|k_y|=|k_z|$. 
	For $\lambda=\frac{\pi}{2}$ the Hamiltonian reduces to a perfect $\bsk. \bsS$ form, where the $S_i$ are a spin-1 representation of the generators of $SO(3)$ group. Indeed at this particular situation, the triple fermions are the spin-1 generalization of the Weyl fermions~\cite{Bradlyn,Bernevig}. 
	The two dimensional analogue of such Hamiltonian with pseudo spin 
	$S=1$ is found in dice lattice~\cite{dice11,dice12,dice13}. 
	
	Let us present the details of the spectrum for $\lambda=\pi/6$. At this particular point the expressions of
	Appendix~\ref{eigen.sec} for the eigenvalues and eigenvectors reduces to,
	$\varepsilon_\pm=\pm |\bf{k}|$ and $\varepsilon_0=\cal O(|\bf{k}^2|)$. 
	Schematic representation of triple bands are illustrated in Fig.~\ref{fig:ek} where the flat band and Dirac cone altogether compose the triple fermions. The circles in Figs.~\ref{fig:ek}.b and ~\ref{fig:ek}.c are the position of 
	chemical potential.  According to the dispersion relation equation, three bands are non-degenerate at $k\neq0$.
	In spherical coordinates where,
	\begin{align}
		{(k_x,k_y,k_z)=|\bf{k}|(\sin\theta \cos\phi,\sin\theta \sin\phi,\cos\theta)},
	\end{align}
	the corresponding eigenfunctions will be represented as,
	\begin{align}
		&\psi_\pm=\frac{1}{\sqrt{2}}
		\begin{pmatrix}
			\pm \sin\theta\\ 
			(\cos\phi\pm i\cos\theta\sin\phi)e^{-i\pi/6}\\ 
			(\mp i\cos\theta\cos\phi+\sin\phi)e^{i\pi/6}
		\end{pmatrix}, \nn\\
		&\psi_0=
		\begin{pmatrix}
			i\cos\phi\\ 
			\sin\theta\sin\phi e^{-i\pi/6}\\ 
			-\sin\theta\cos\phi e^{i\pi/6}
		\end{pmatrix}.
		\label{eq.3}
	\end{align}
	The Chern number arising from the above band touching point for each of the above bands 
	are $\nu_\pm=\pm2$ and $\nu_0=0$. 
	The flat middle band in Fig.~\ref{fig:ek} is where the Chern number changes sign across which the anomalous Hall effect changes sign~\cite{chamon}. 
	
	\begin{figure}[t]
		\centering
		\includegraphics[width=0.75\linewidth]{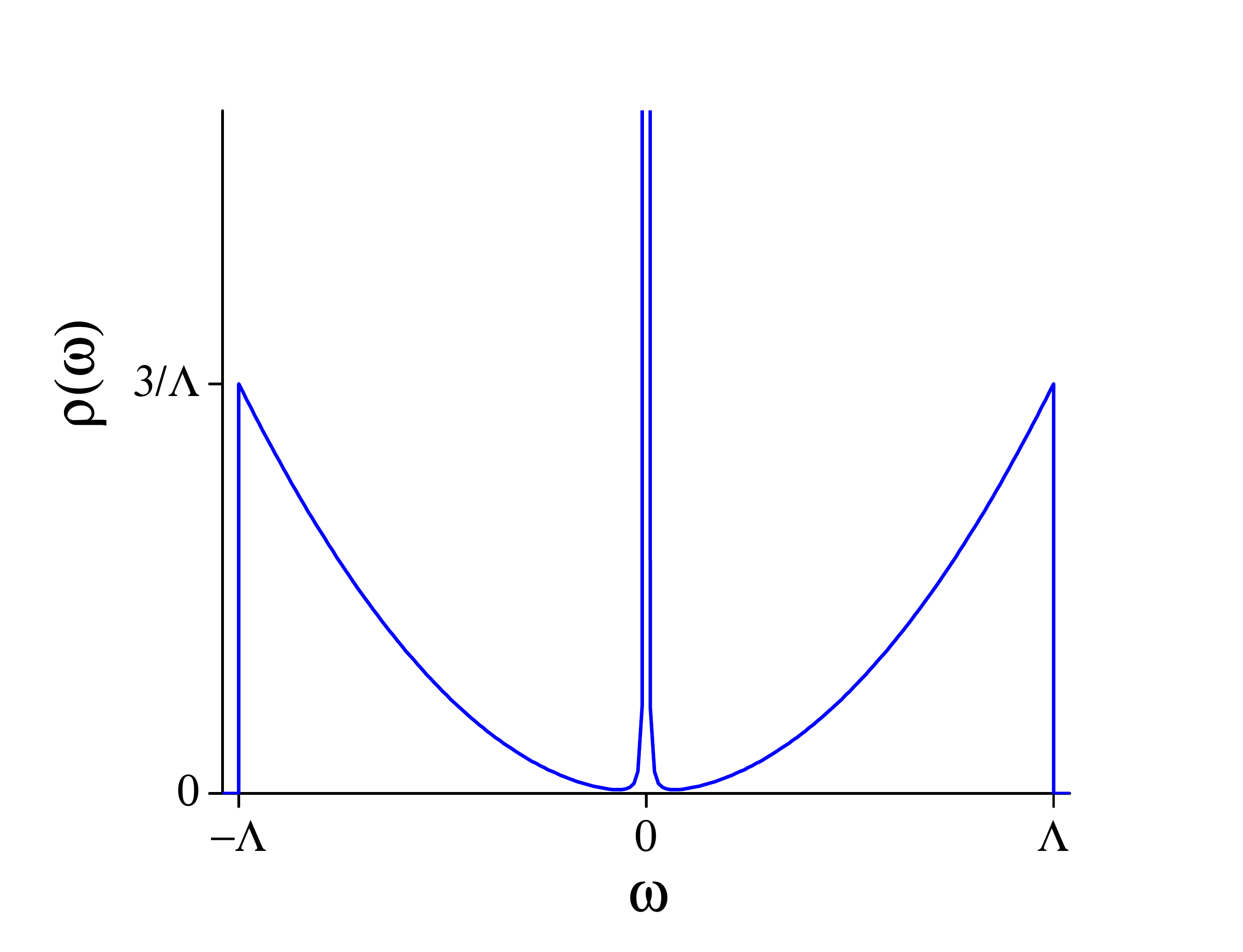}
		\caption{(Color online)  Density of states of the triple fermions in terms of $\omega$. The $\Lambda $ is the energy cutoff.}
		\label{fig:dos}
	\end{figure}
	
	The Green function for this effective Hamiltonian can be calculated as follows. Starting from the
	general definition
	\begin{eqnarray}
		G({\bf{k}},i\omega ) =1/(\omega+ i\eta-\hat H(\bf{k})),
	\end{eqnarray} 
	where $i\omega=\omega+i\eta$ is Matsubara frequency for our case it becomes,
	\begin{align}
		&\hat G({\bf{k}},i\omega ) = \frac{1}{{i\omega \left( {i{\omega ^2} - {\bf{k}^2}} \right)}}\times\nonumber\\
		& \left( {\begin{array}{*{20}{c}}
				{i{\omega ^2} - {k_z}^2}&{\nu^*}^2{k_y}{k_z} + \nu{k_x}i\omega &\nu^2{k_x}{k_z} + \nu^*{k_y}i\omega\\
				\nu^2{k_y}{k_z} +\nu^*{k_x}i\omega &{i{\omega ^2} - {k_y}^2}&{\nu^*}^2{k_x}{k_y} + \nu{k_z}i\omega \\
				{\nu^*}^2{k_x}{k_z} +\nu{k_y}i\omega &\nu^2{k_x}{k_y} +\nu^*{k_z}i\omega &{i{\omega ^2} - {k_x}^2}
		\end{array}} \right)
	\end{align}
	where as pointed out earlier since $|a|=1$, $a=e^{i\lambda}$ will be a pure phase. 
	At the particular point $\lambda=\pi/6$ we denote $a$ by $\nu = \rm{e}^{\frac{i\pi}{6}}$. 
	From the Green's function above, the spectral function will be,
	\begin{eqnarray}
		\rho ({\bf{k}},i\omega) =-\frac{1}{\pi}\lim_{\eta \to 0}\text{Im} \text{Tr}  \hat{G}({\bf{k}},i\omega\to\omega+i\eta)
	\end{eqnarray}
	which simplifies to,
	\begin{align}
		\rho({\bf{k}},i\omega)=-\frac{1}{\pi}\lim_{\eta \to 0}\text{Im} 
		\frac{{3{i\omega ^2} - {{\bf{k}}^2}}}{{i\omega \left( {{i\omega ^2} - {{\bf{k}}^2}} \right)}}.
	\end{align}
	The total density of states (DOS) is therefore obtained as, 
	\begin{align}
		\rho(\omega)&=\frac{1}{\frac{4}{3}\pi \Lambda^3}\int_{0}^{\Lambda}{4\pi k^2dk\rho(\bf{k},\omega)} \nonumber \\
		&=\delta(\omega)+\frac{3\omega^2 \Theta\left(\Lambda-|\omega|\right)}{\Lambda^3}.
	\end{align}
	Here the $\Lambda$ is the energy cutoff and $\Theta$ is Heaviside step function. The total DOS is plotted in Fig.~\ref{fig:dos}, and due to the 
	triple bands in the system the integration of DOS gives $3$. This can be used in any integration involving the
	summation over states needed in calculation of thermodynamic quantities. 
	For example, the filling factor defined by,
	\begin{eqnarray}
		n(\mu)=\int_{-\infty}^{\mu}d\omega\rho(\omega).
	\end{eqnarray}
	will become,
	\begin{align}
		n(\mu)=\Theta(\mu)+\left[1+\left(\frac{\mu}{\Lambda}\right)^3\right]\Theta\left(\Lambda-|\mu|\right).
	\end{align}
	
	\begin{figure}[b]
		\centering
		\includegraphics[width=0.7\linewidth]{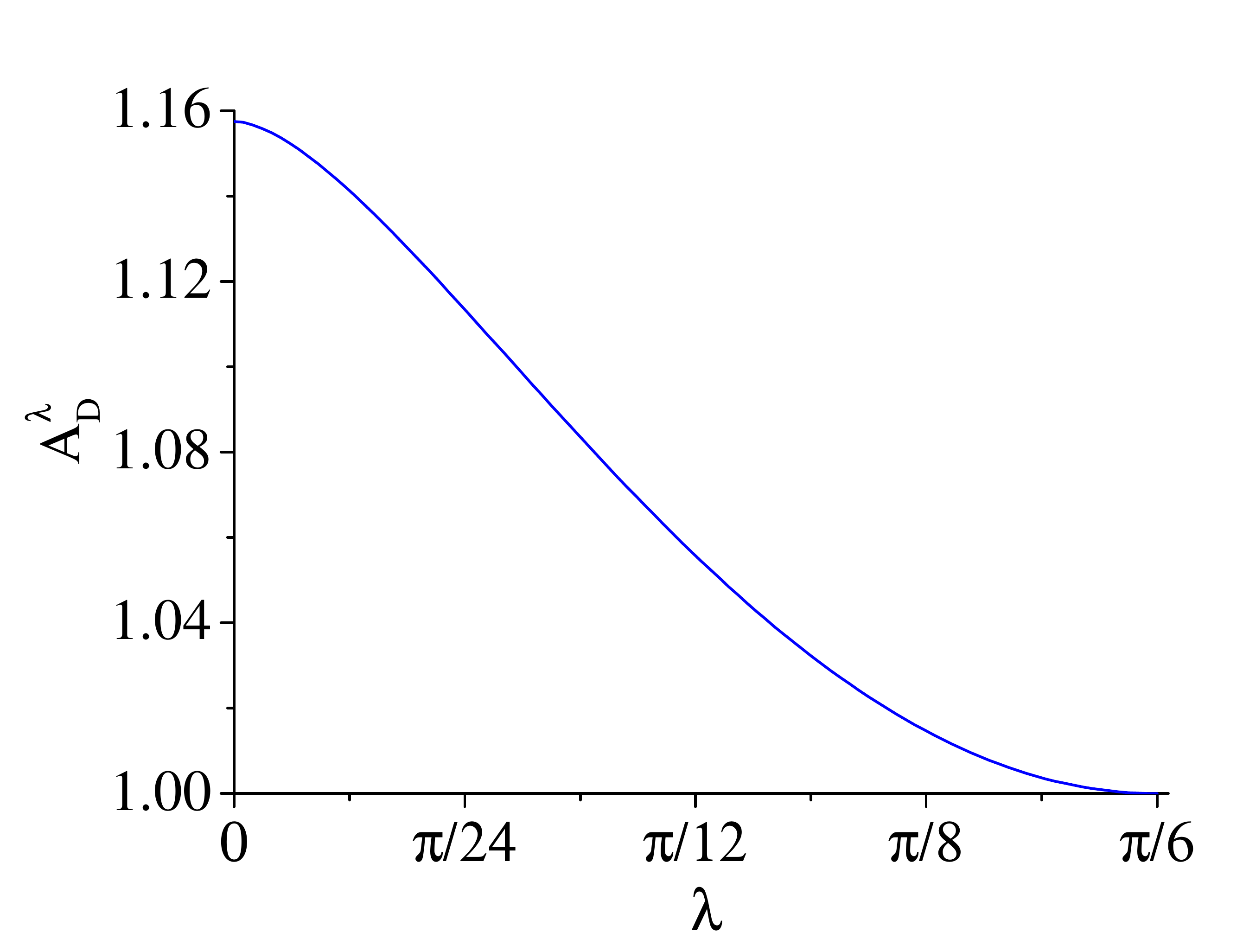}
		\caption{(Color online)  Dependence of the numerical weight $A^\lambda_D$ of the Drude weight on the phase $\lambda$. in Eq.~\eqref{drude2.eqn}.}
		\label{fig:drudeweight}
	\end{figure}

	\section{optical conductivity}\label{op1}
	In this section, we are going to calculate the optical conductivity by considering the three-fold degeneracy. 
	The intraband component which gives the Drude contribution and interband component of the optical conductivity is considered. 
	Our calculations are based on the Kubo formula. 
	Working in the one-loop approximations, the real part of the Kubo 
	formula at finite frequency is~\cite{carbotte2}, 
	\begin{align}
		\sigma_{\alpha\beta}(\Omega)=\frac{\sigma_0\pi}{\Omega}&\int_{-\infty}^{\infty}d\omega[f(\omega)-f(\omega+\Omega)]\\
		\times& \int\frac{d^3\bsk}{(2\pi)^3}\text{Tr}[\hat v_\alpha \hat A(\bsk,\omega)\hat v_\beta \hat A(\bsk,\omega+\Omega)],\nn
	\end{align}
	where $f(\omega)=[\exp(\beta(\omega-\mu))+1]^{-1}$ is the Fermi function. $\beta$ 
	is $1/{k_BT}$ and  $\mu$ is the chemical potential and $\alpha,\beta$ run over Cartesian indices $x,y,z$. 
	The velocity operator $\hat v_\alpha$ is related to the Hamiltonian via $\hat v_\alpha=\partial \hat H / \partial k_\alpha$. 
	The spectral function $\hat A$ and the Green's function are related by,
	\begin{align}
		\hat{G}(\bsk,z)=\int_{-\infty}^{\infty}\frac{\hat A(\bsk,\omega)}{z-\omega}d\omega
	\end{align}
	where
	\begin{align}
		\hat{G}^{-1}(\bsk,z)=\hat I z-\hat H
	\end{align}
	and $\hat I$ is identity matrix. 
	
	\subsection{Intraband contribution: Drude weight}
	After some calculations (see Appendix~\ref{oc.sec}) the intraband component of 
	the optical conductivity gives the following Drude weight $W_D$ ($\text{lim}\Lambda\rightarrow \infty$), 
	\begin{align}
		\sigma_{\alpha\beta}^D(\Omega)=\frac{ \sigma_0}{6\pi}\left(\mu^2+\frac{\pi^2}{3}k_B^2T^2\right)\delta(\Omega) \equiv W_D\delta(\Omega)
		\label{drude1.eqn}
	\end{align}
	where $T$ is the temperature and we assume $\mu\ne 0$. 
	The $1/(6\pi)$ factor is similar to the case of Dirac/Weyl materials.
	The $T^2$ dependence and its $\pi^2/3$ coefficient is a characteristic of free 
	fermions~\cite{carbotte,carbotte2} and comes from a Sommerfeld expansion. 
	This coefficient is related to the number of degrees of freedom or central charge in conformally invariant theories. 
	For non-zero $\mu$, even at $T=0$, the Drude weight does not vanish.
	
	For $\lambda\ne \pi/6$ we obtain Drude weight by employing the numerical calculations which gives (see Fig.~\ref{fig:drudeweight}),
	\begin{align}
		W_D^\lambda=A_D^\lambda\frac{\sigma_0}{6\pi}\left(\mu^2+\frac{\pi^2}{3}k_B^2T^2\right)
		\label{drude2.eqn}
	\end{align}
	where $A_D^\lambda$ is a numeric coefficient obtained numerically and plotted in Fig.~\ref{fig:drudeweight}.  
	As shown in this figure, the exact value of $A_D^\lambda$ for $\lambda=\pi/6$ is $1$  in agreement with the 
	analytic result in Eq.~\eqref{drude1.eqn}. 
	Depending on the value of $\lambda$, there can be about $16$ percent variations in the Drude weight. This variation can be employed as an optical determination of the phase $\lambda$. 
	The physical significance of such a dependence of the Drude weight to the phase $\lambda$ is that
	a simple Drude weight measurement enables us to map the value of the phase $\lambda$ in $a=e^{i\lambda}$ of Eq.~\eqref{kdotS.eqn}.
	As we will see shortly, the dependence on the phase $\lambda$ will also show up in interband processes as well.

	\begin{figure}[t]
		\centering
		\includegraphics[width=1.00\linewidth]{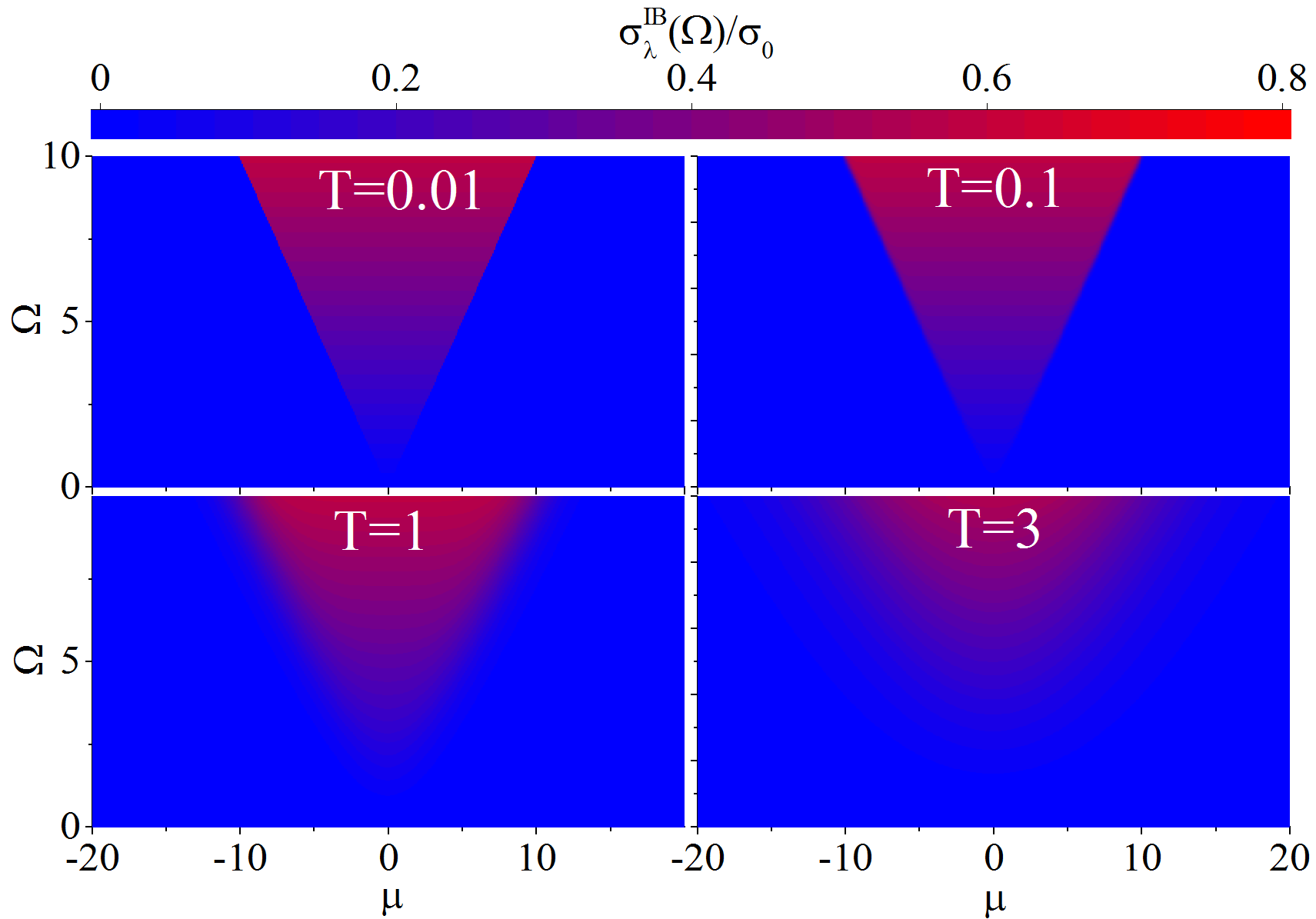}
		\caption{(Color online) Intensity map of the interband optical conductivity of triple fermions as a function of $\mu$ and $\Omega$ for different values of temperature at $\lambda=\pi/6$.
			The temperature is in units of $\hbar v_F|k|/K_B$. }.
		\label{fig:conductivity}
	\end{figure}
	
	\begin{figure}[t]
		\centering
		\includegraphics[width=1.00\linewidth]{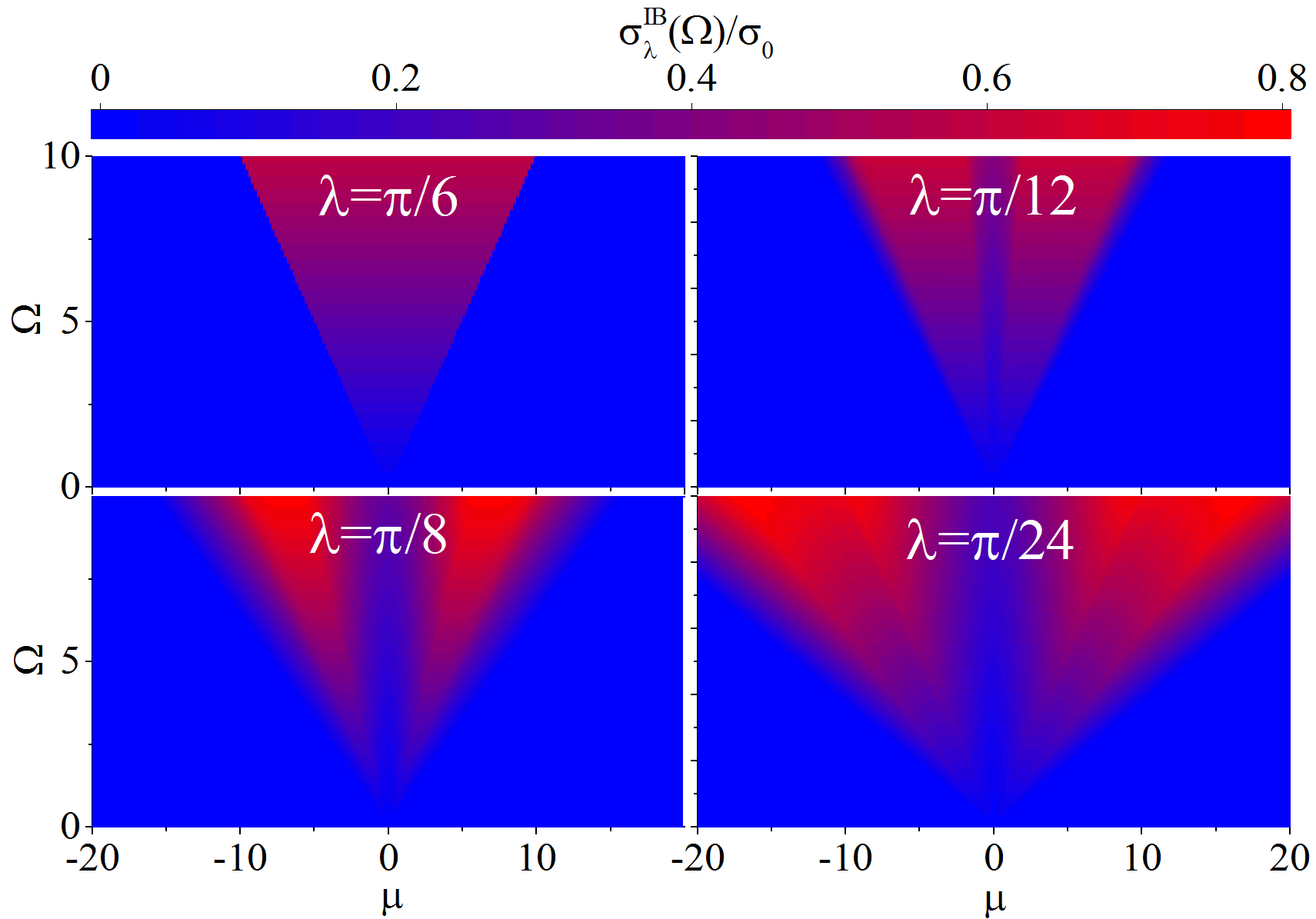}
		\caption{(Color online) Intensity map of the interband optical conductivity of triple fermions 
			as a function of $\mu$ and $\Omega$ for different values of $\lambda$ at $T=0.01$. 
		}
		\label{fig:sigma-Lam}
	\end{figure}
	
	\subsection{Interband contribution}
	The interband conductivity of three-fold degenerate states is also sensitive to the phase $\lambda$. 
	As detailed in appendix~\ref{oc.sec}, at $\lambda=\pi/6$ it will be given by,
	\begin{align}
		\sigma_{xx}^{\text{IB}}(\Omega)    = \frac{\sigma_0}{6\pi}\Omega\left(f(-\Omega)-f(\Omega)\right)
	\end{align} 
	The above conductivity in at $\mu=0$ is,
	\begin{align}
		\sigma_{xx}^{\text{IB}}(\Omega)    = \frac{\sigma_0}{6\pi}\Omega\tanh{\frac{\beta \Omega}{2}}.
	\end{align} 
	For  $|\beta\Omega|\ll1$ it reduces to,
	\begin{align}
		\sigma_{xx}^{\text{IB}}(\Omega)    \propto
		\frac{\sigma_0}{12\pi}\frac{\Omega^2}{k_BT}
		\label{xx1}
	\end{align} 
	while for $|\beta\Omega|\gg1$  is will be given by,
	\begin{align}
		\sigma_{xx}^{\text{IB}}(\Omega)    \propto
		\frac{\sigma_0}{6\pi}\Omega
		\label{xx2}
	\end{align} 
	This helps us to understand Fig.~\ref{fig:conductivity} which is an intensity plot of the interband optical absorption for $\lambda=\pi/6$. In this figure and subsequent figures,
	units of temperature, $\Omega$ and $\mu$ are renormalized to $\hbar v_f$. The salient feature of these figures is a wedge-shaped absorption edge which is more manifest for the lowest temperature $T=0.01$     which is displayed in the top left panel. To understand the wedge, if we walk along a constant $\Omega$ line,
	increasing $\mu$ causes more and more Pauli blocking for optical absorption and hence beyond a certain point, the optical absorption is diminished by Pauli blocking. Therefore this wedge can be called a Pauli blockade wedge. Upon increasing the temperature, the Pauli blockage wedge will be less sharped. The tip of the wedge at $\mu=0$ at high temperatures is given by Eq.~\eqref{xx1} which is inversely proportional to temperature and therefore for higher temperature will tend to zero as the inverse of temperature. 
	This description explains why the tip of the wedge turns blue in the bottom right panel. 
	
	In Fig.~\ref{fig:sigma-Lam} the optical absorption is numerically mapped for a fixed temperature $T=0.01$. Upon getting away from $\lambda=\pi/6$, first of all, the Pauli blockade wedge becomes wider. This broadening is most manifest in the bottom right panel. The intensity is also increased by moving away from $\lambda=\pi/6$ which can be seen as an enhanced red contribution in color scheme. 
	Another effect of moving away from $\lambda=\pi/6$ is that the optical absorption around the $\mu=0$ is depleted. This reduction has to do with the strong anisotropy of the eigenvalue of the triple fermions which will be explained in the sequel.  
	
	\begin{figure}[t]
		\centering
		\includegraphics[width=1\linewidth]{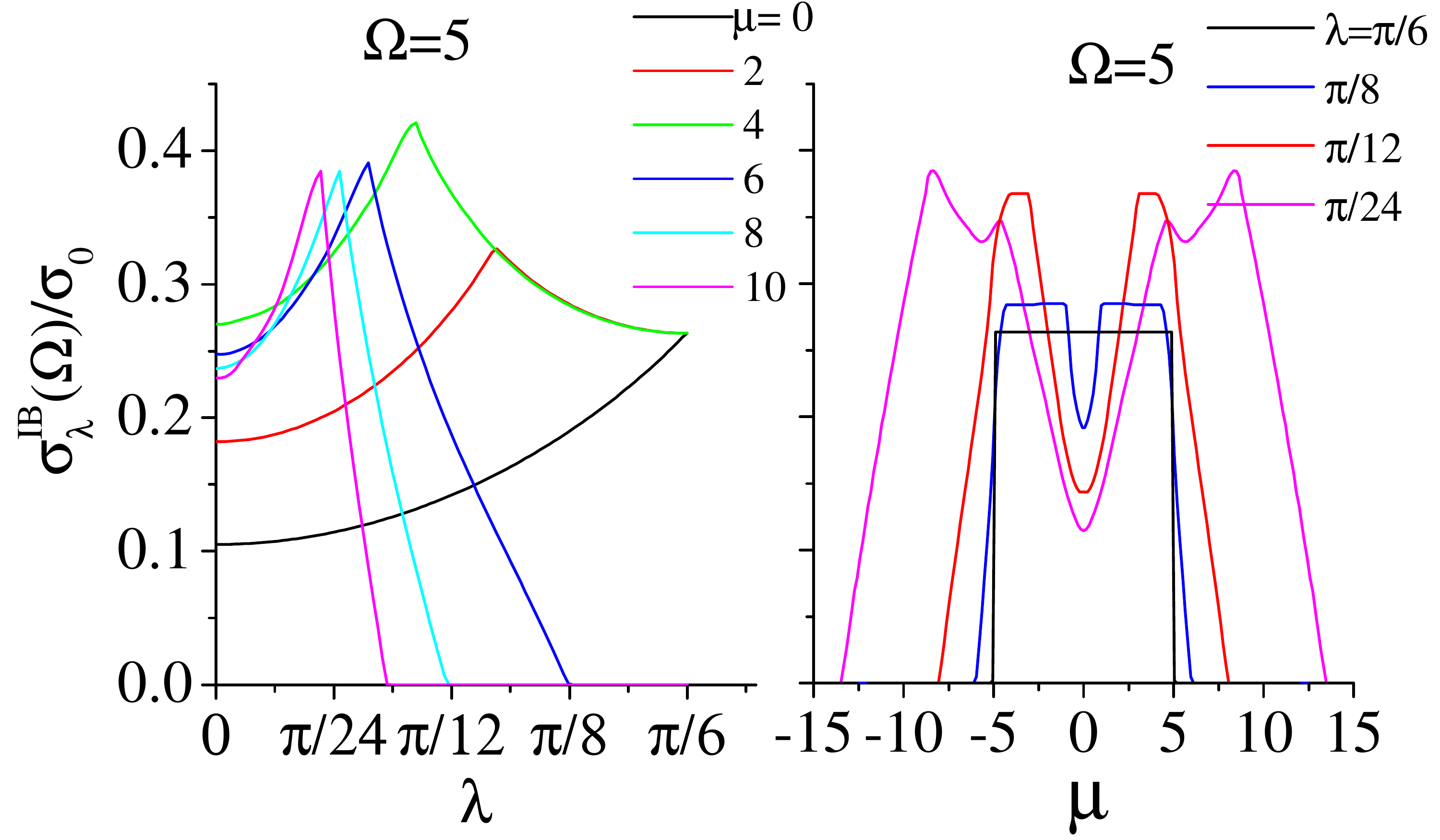}
		\caption{
			(Color online) Optical absorption at fixed temperature $T=0.01$ and $\Omega=5$ in natural
			unit $\sigma_0=e^2/\hbar$.
			Left: The $\lambda$-dependence for a constant $\Omega$ for various values of $\mu$. 
			Right: $\mu$-dependence for various values of $\lambda$. 
		}
		\label{fig:2D}
	\end{figure}
	
	To learn more about the optical absorption of this system, let us present cuts Figs.~\ref{fig:conductivity} and~\ref{fig:sigma-Lam}. 
	In Fig.~\ref{fig:2D} a constant $\Omega=5$ cut for a constant $T=0.01$ temperature has been plotted. In the left panel, we plot the $\lambda$ dependence of the optical absorption for various values of $\mu$. In the right panel, we plot the chemical potential dependence of the optical absorption for various values of $\lambda$. The $\lambda$ dependence has a cusp that moves to lower $\lambda$ by increasing $\mu$. 
	For every value of $\Omega$ and $\mu$, the strong $\lambda$ dependence can be very helpful in the optical determination of $\lambda$. 
	The $\mu$ dependence is more interesting. Both flat part and sharp drop of the $\mu$ dependence in $\lambda=\pi/6$ point will be affected by moving away from $\lambda=\pi/6$. First, a depletion around $\mu=0$ is developed which is accompanied by enhancement of the absorption in two sides of the depletion region. Second, the sharp drop at the Pauli blockade wedge is replaced by a smoother drop. 
	
	\begin{figure}
		\centering
		\includegraphics[width=1\linewidth]{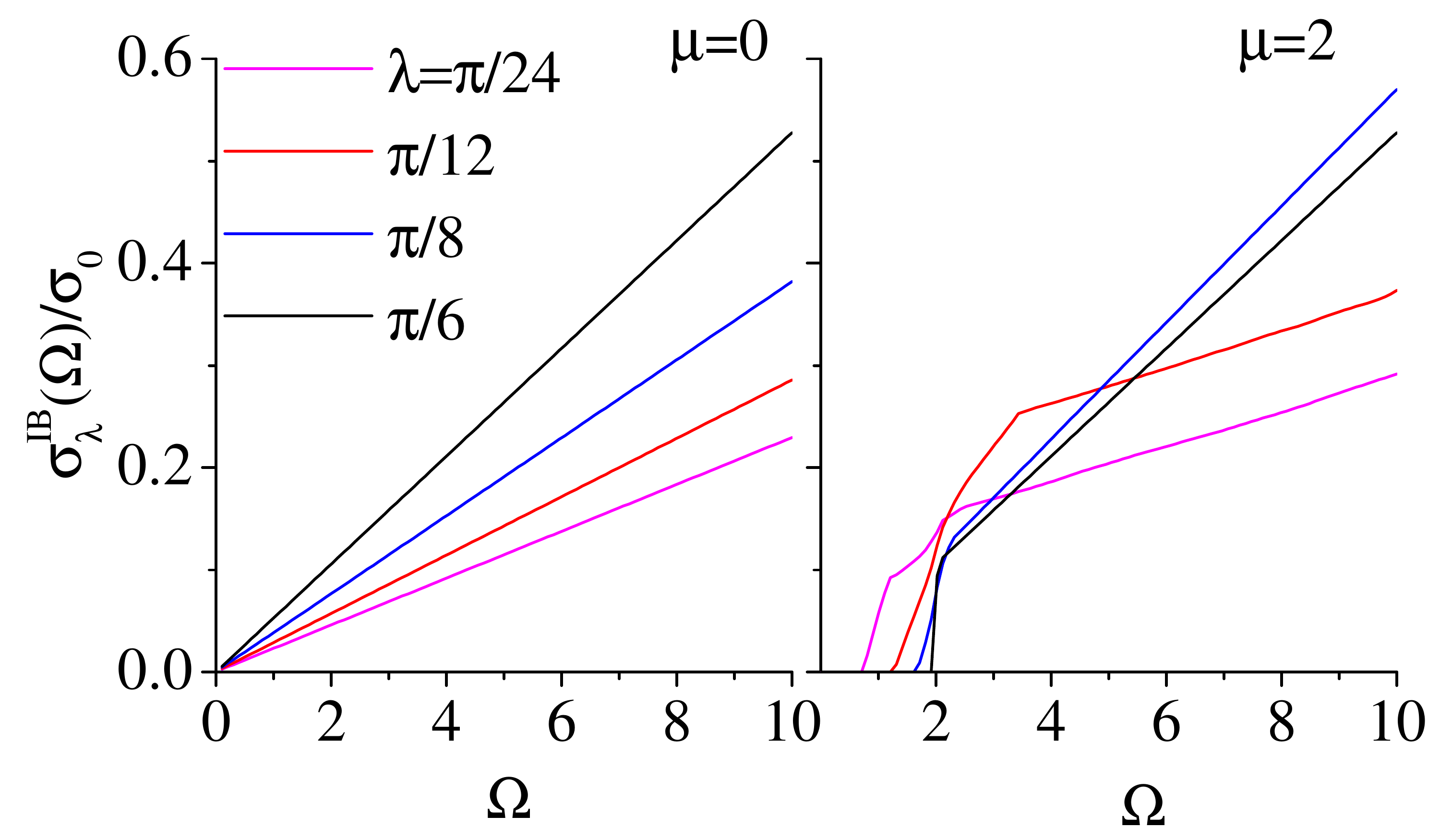}
		\caption{(Color online) Energy dependence of the optical absorption at $T=0.01$ for $\mu=0$ (left) and $\mu=2$ (right) and
			various values of $\lambda$. }
		\label{fig:sigma-omega}
	\end{figure}
	
	Fig.~\ref{fig:sigma-omega} shows the energy dependence of the optical conductivity of the system at low temperature, $T=0.01$. The left panel is for $\mu=0$ while the right panel is for $\mu=2$. Different colors correspond to various values of $\lambda$ as indicated in the legend.
	Again a very strong $\lambda$ dependence can be seen in the optical absorption lineshapes. First of all for $\mu=0$ in the left panel the lineshape is linear in $\Omega$. This is in agreement with Eq.~\eqref{xx2}. Note that for very small $\Omega$ satisfying $|\Omega \beta|\ll 1$,
	Eq.~\eqref{xx1} the optical absorption will depend quadratically on $\Omega$ which crosses over to the linear dependence in Eq.~\eqref{xx2}.
	The left panel of Fig.~\ref{fig:sigma-omega} clearly displays this linear dependence. As can be seen, the slope of the absorption lineshape is clearly controlled by the phase factor $\lambda$ and increases by going away from $\lambda=\pi/6$ point. 
	For $\mu=2$ (right panel), as long as $\Omega$ is large enough compared to $\mu$, still, a linear dependence on $\Omega$ can be seen.
	In addition to that, the behavior of optical absorption near the Pauli blockade wedge (onset of absorption in the right panel) also depends on the phase $\lambda$. 
	
	\subsection{"Helicity" reversal transitions}
	\begin{figure}[b]
		\centering
		\includegraphics[width=0.650\linewidth]{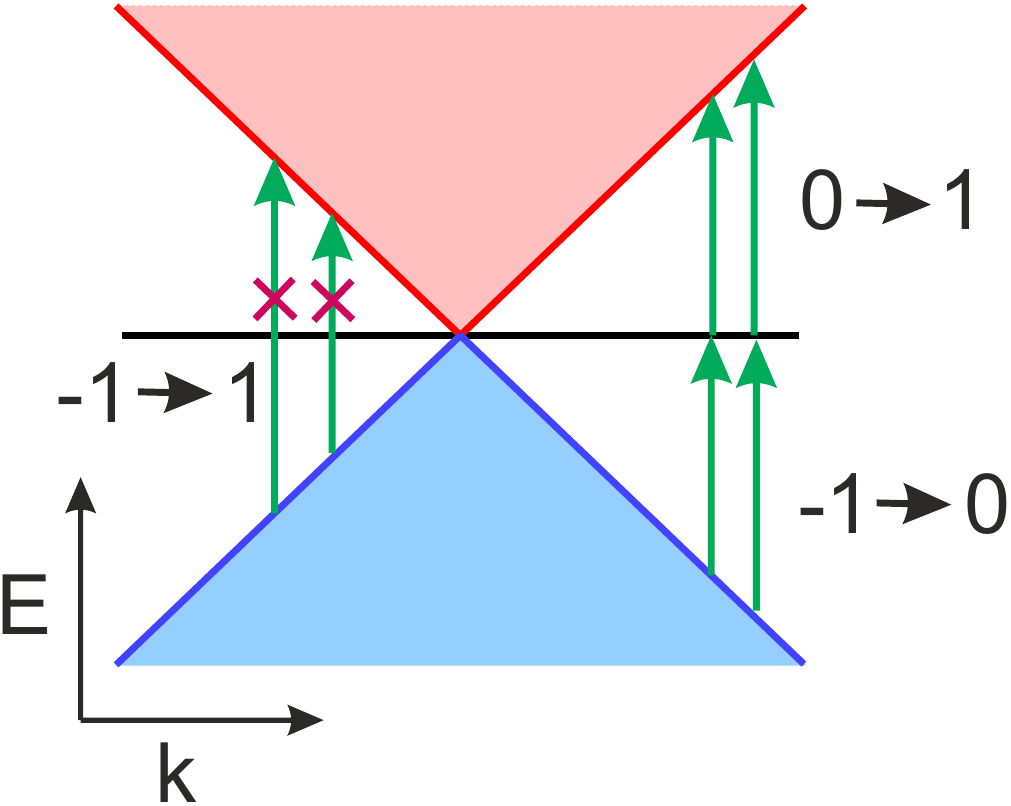}
		\caption{(Color online)  Structure of dipole matrix elements in triple point Fermion systems at $\lambda=\pi/6$. 
			At this point the helicity reversal transitions $m=-1\to m=+1$ transitions are not allowed.
		}
		\label{band}
	\end{figure}
	As can be seen from Eq.~\eqref{eq.3}, the three bands are labeled with $m=0,\pm 1$. In the $\lambda=\pi/2$ point where the Hamiltonian
	becomes $\bsk.\bsS$ the band index $m$ can be interpreted as the spin angular momentum along $\bsk$. This is similar to the "helicity" $\bsk.{\boldsymbol \sigma}$
	of ordinary fermions. Let us continue to use the name "helicity" for the band index $m$ for generic values of $\lambda$. 
	Closely examining the optical transition matrix elements between $m=-1$ and $m=+1$ (see Fig.~\ref{band}) shows that the
	dipole matrix elements for these transitions that reverse the sign of $m$ are exactly zero for $\lambda=\pi/6$. This phenomena can be directly seen by using Eq.~\eqref{eq.3} to form the velocity matrix element between the eigen-vectors $m=\pm1$. 
	This has been illustrated in Fig.~\ref{band}.

	\begin{figure}[t]
		\centering
		\includegraphics[width=1\linewidth]{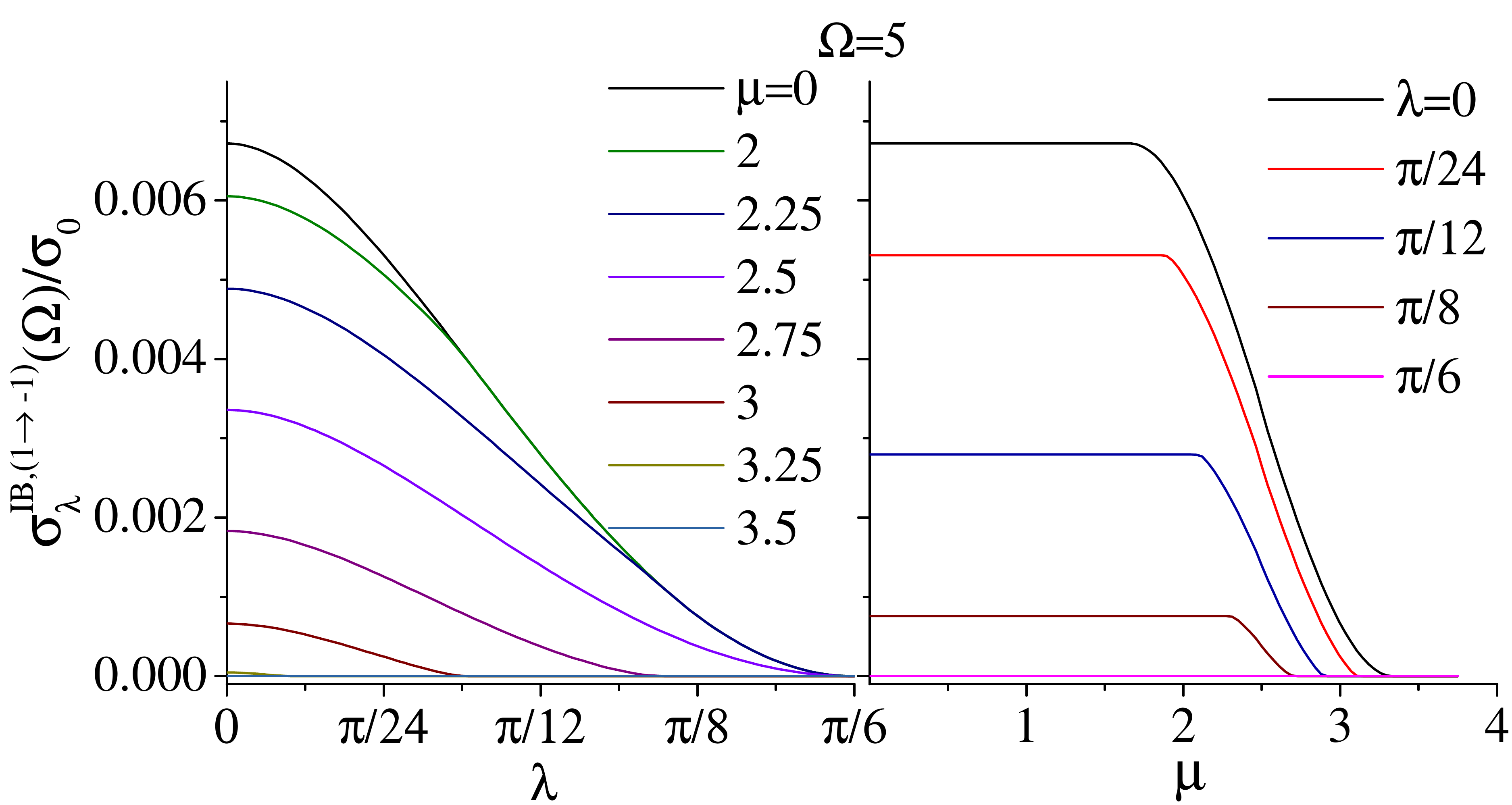}
		\caption{(Color online) Helicity reversal interband transitions $m=-1\to m=+1$ in triple point fermions for fixed $T=0.01$ and $\Omega=5$.
			(left) The $\lambda$ dependence for various values of $\mu$. As can be seen for all values of $\mu$, the partial optical absorption arising
			from $m=-1\to m=+1$ is zero in a range of $\lambda$ values around $\lambda=\pi/6$. 
			(right) The $\mu$ dependence of the partial optical absorption for various values of $\lambda$.
		}
		\label{fig:sigma-lambda-interband}
	\end{figure}
	
	By moving away from $\lambda=\pi/6$ the vanishing dipole matrix elements for $m=-1\to m=+1$ transitions 
	continuously start to pick non-zero values. To quantify these processes, in Fig.~\ref{fig:sigma-lambda-interband}
	we have plotted the contribution of these processes in optical absorption for a fixed low temperature $T=0.01$ and a typical energy $\Omega=5$. As can be seen in the left panel, the partial optical absorption is exactly zero at $\lambda=\pi/6$ for all values of $\mu$. This phenomena is in agreement with the schematic sketch in Fig.~\ref{band}. 
	Then it continuously rises from zero by deviating from $\lambda=\pi/6$ point.
	This rise from zero is slower for larger chemical potentials. The same partial optical absorption is plotted in the right panel as a function of $\mu$ for various values of $\lambda$ indicated in the legend. Again for $\lambda=\pi/6$ for the entire range of $\mu$ values the partial optical absorption is zero which is in agreement with Fig.~\ref{band}. By deviating from $\lambda=\pi/6$, the low $\mu$ part gradually rises which indicates that the "helicity" reversal transitions are gaining weight. The large $\mu$ portions of the partial absorption are also diminished by approaching the Pauli blockade scale $\sim\Omega$. 
	
\begin{figure}[t]
		\centering
		\includegraphics[width=1\linewidth]{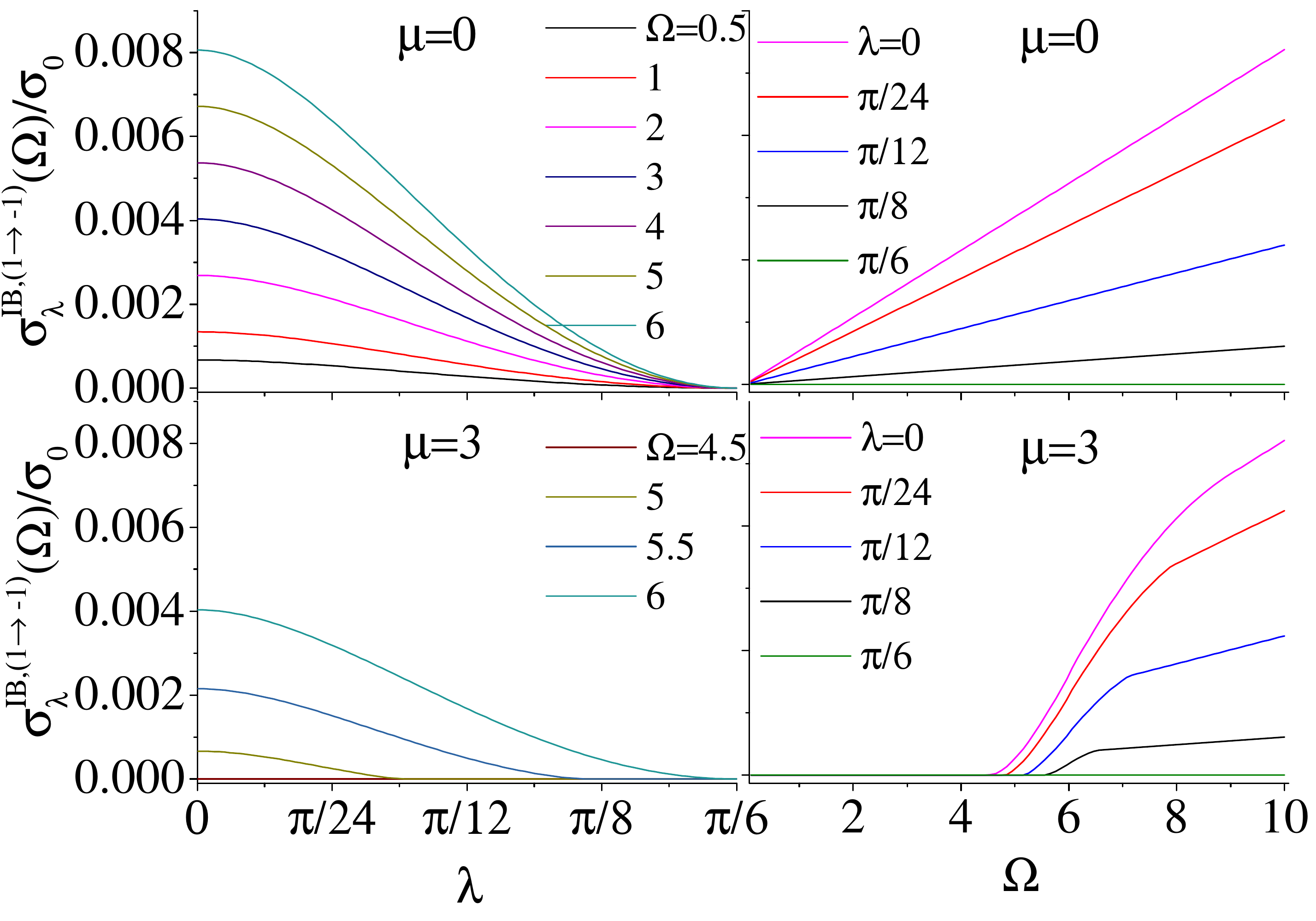}
		\caption{ (Color online)  Frequency dependence (right column) and $\lambda$-dependence (left column) of the
			interband portion of the optical conductivity.
		}
		\label{fig10.fig}
	\end{figure}

	In the left column of Fig.~\ref{fig10.fig} we have plotted the $\lambda$ dependence of the interband portion of the optical conductivity for $\mu=0$ (upper panel) and $\mu=3$ (lower panel) for indicated values of $\Omega$. Again it can be seen that at $\lambda=\pi/6$, there is no helicity reversal absorption. By moving away from this point, the $\mu=0$ panel shows a smooth increase of the absorption for all frequencies. In the lower left panel corresponding to $\mu=3$, for those frequencies less than $\sim \mu$ the absorption is zero due to Pauli blockade. For larger frequencies, the possible non-zero helicity reversal optical absorption can only take place away from $\lambda=\pi/6$. 
	
	The right column in this figure shows the helicity reversal $m=-1\to m=+1$ absorption lineshape for
	$\mu=0$ (upper panel) and $\mu=3$ (lower panel). For $\mu=0$ there is no Pauli blockade, and for all indicated values of $\lambda$, the $\Omega$-dependence is linear. For $\lambda=\pi/6$, the line is horizontal and corresponds to zero absorption. For other values of $\lambda$, the slope of the line is non-zero and increases by moving away from $\lambda=\pi/6$. In the $\mu=3$ case (lower right panel), 
	there is a region of Pauli blockade where the helicity reversal part of the optical absorption is zero for all values of $\lambda$. For higher frequencies, the $\lambda=\pi/6$ lineshape continues to remain zero, while for the other values of $\lambda$, the helicity reversal transitions become possible. 
	
	\section{Discussions and concluding remarks} \label{results}
\begin{figure}[t]
		\centering
		\includegraphics[width=1\linewidth]{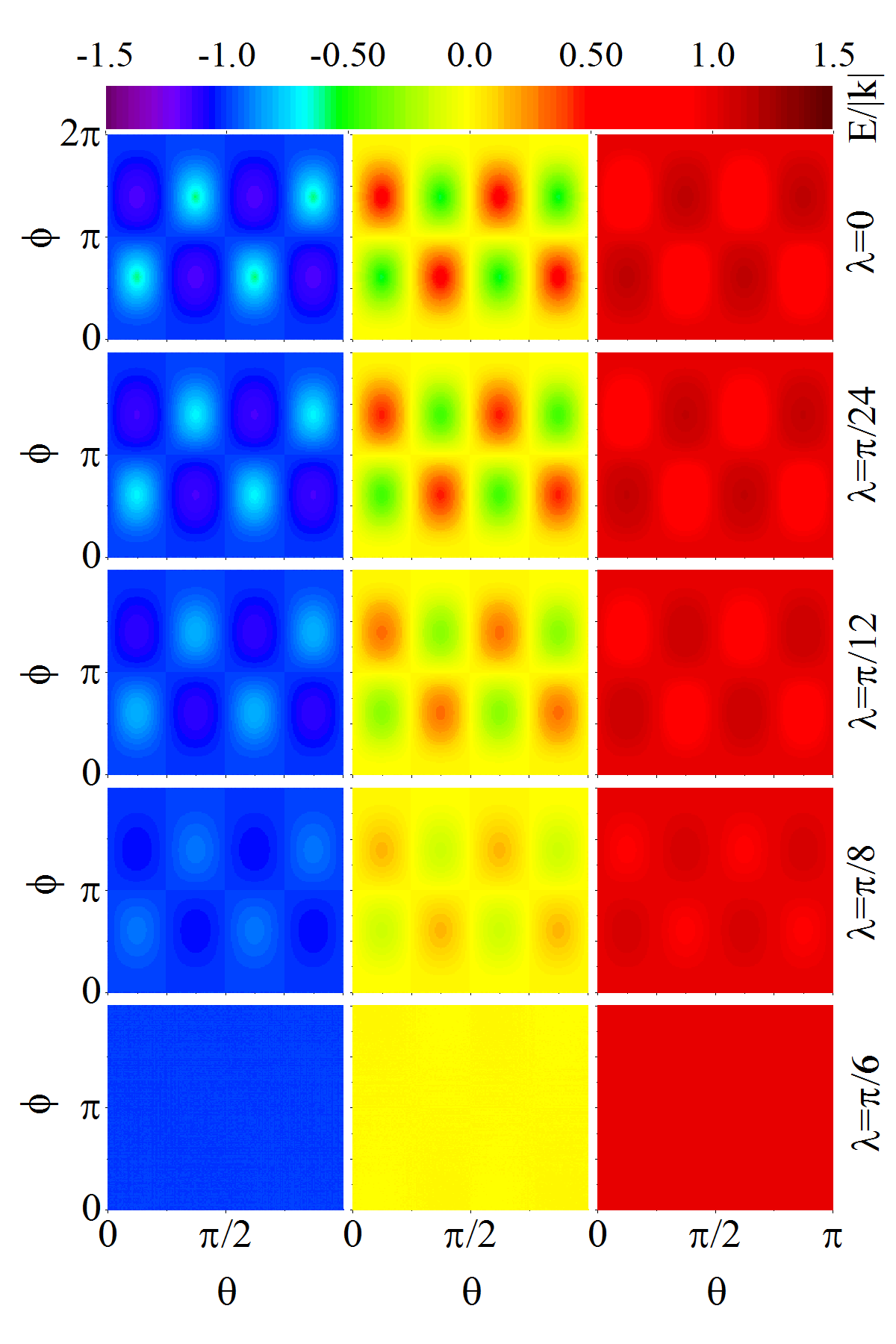}
		\caption{ (Color online)  Dispersion $E/|k|$ of triple fermions in terms of $(\theta, \phi)$. As illustrated, the blue, yellow and red regions respectively are belong to $-1$, $0$ and $1$ values of energy. By tunning the $\lambda$ in terms of $(\theta, \phi)$ they are not uniform and shows some fluctuations. }
		\label{fig:graph10}
	\end{figure}

	As noted in Figs.~\ref{fig:conductivity},~\ref{fig:sigma-Lam} and~\ref{fig:2D} the $\mu$ dependence 
	is non-trivial in the sense that $\mu$ does not set a sharp step-function-like edge for the Pauli blockade. 
	This has to do with the anisotropy of the band structure. 
	To obtain the total optical absorption, one must integrate over the states at wave-vector $\bsk$
	which has been converted to an integral over $\omega$ of two terms 
	$P_1^{\text{IB}}(\omega,\Omega)$ and $P_2^{\text{IB}}(\omega,\Omega)$ (see appendix \ref{oc.sec}).
	To examine the role each wave vector $\bsk$, in Fig.~\ref{fig:graph10} we have plotted the dependence of the
	energy dispersion $E_m(\bsk)$ on the solid angle $(\theta, \phi)$ on a sphere with $|\bsk|=1$
	for $m=-1$ (left column), $m=0$ (middle column) and $m=+1$ (right column). 
	Each row corresponds to the value of $\lambda$ indicated in the legend. 
	As can be seen in the bottom row corresponding to $\lambda=\pi/6$ the solid angle dependence is trivial and, everything is isotropic. By deviating $\lambda$ from $\pi/6$, non trivial angular dependence
	in all bands $m=0,\pm 1$ arises. Therefore for a given $\mu$ some directions of the pertinent band sink below $\mu$, thereby creating "puddles" of filled states and hence making a phase space for optical absorption and some states float above $\mu$. 
	This anisotropy is behind a non-trivial dependence of the optical absorption to the chemical potential. 
	These puddles are responsible for the non-vanishing $m=-1\to m=+1$ transitions for $\lambda\ne \pi/6$ in Fig.~\ref{fig:sigma-lambda-interband} and
	Fig.~\ref{fig10.fig}. 
	
	In conclusion as an example of the effective theory of non-symmorphic lattices, we have considered a generic three-fold 
	degenerate Hamiltonian which describes triple fermions parameterized by a phase $\lambda$. 
	The parameter $\lambda$ controls both the Drude and interband part of the optical absorption.
	This can be employed as an optical method for the experimental determination of the angular parameter $\lambda$. 
	In particular, at the fine-tuned value of $\lambda=\pi/6$ (which provides an analytical handle), the dipole matrix elements for $m=-1\to m=+1$ transitions are identically zero and rise by moving away from this point. 
	
	Not surprisingly, the $T^2$ and $\mu^2$ dependence of the Drude weight being a property of free fermions also holds in triple point Fermions. The interesting point, however, is that the deviation from $\lambda=\pi/6$
	renormalizes the overall numerical coefficient with respect to the corresponding coefficients in Dirac/Weyl systems (Fig.~\ref{fig:drudeweight}).
	The deviation of the $T^2$ coefficient from its universally Dirac/Weyl value encoded in number $A^\lambda_D$ of Eq.~\eqref{drude2.eqn}
	serves as quick optical discrimination of Dirac/Weyl degeneracies from triple point degeneracies. 
	
	\section{acknowledgements}
	T.F. appreciates the financial support from Iran National Science Foundation (INSF) under postdoctoral project No. 96015597. S.A.J. appreciates Iran Science Elites Federation (ISEF) and research deputy of Sharif University of Technology, grant No. G960214.

	\newpage
	\appendix
		\begin{widetext}

		\section{Eigenvalues and Eigenvectors discussion}
		\label{eigen.sec}
		In this appendix first of all we obtain the eigenvalues and 
		eigenvectors of the general Hamiltonian. Then we describe the details of 
		calculation for obtaining the optical conductivity of the three-fold degenerated fermions. By considering the general form of the Hamiltonian, 
		\begin{align}
			H_{199}=\begin{pmatrix}
				0&  a k_x&a^* k_y  \\ 
				a^* k_x& 0 &  a k_z\\ 
				a k_y& a^* k_z & 0
			\end{pmatrix} 
		\end{align}
		the eigenvalues and eigenvectors are respectively as, 
		\begin{align}
			\omega_+ \to& \frac{\sqrt[3]{\sqrt{729 \sin ^4(\theta ) \cos ^2(\theta ) \cos ^2(3 \lambda ) \sin ^2(2 \phi )-108}+27 \sin ^2(\theta ) \cos (\theta ) \cos (3 \lambda ) \sin (2 \phi )}}{3 \sqrt[3]{2}}
			\\
		&	+\frac{\sqrt[3]{2}}{\sqrt[3]{\sqrt{729 \sin ^4(\theta ) \cos ^2(\theta ) \cos ^2(3 \lambda ) \sin ^2(2 \phi )-108}+27 \sin ^2(\theta ) \cos (\theta ) \cos (3 \lambda ) \sin (2 \phi )}} \nn
			\\
			\omega_- \to& -\frac{\left(1-i \sqrt{3}\right) \sqrt[3]{\sqrt{729 \sin ^4(\theta ) \cos ^2(\theta ) \cos ^2(3 \lambda ) \sin ^2(2 \phi )-108}+27 \sin ^2(\theta ) \cos (\theta ) \cos (3 \lambda ) \sin (2 \phi )}}{6 \sqrt[3]{2}}
			\\
			&-\frac{1+i \sqrt{3}}{2^{2/3} \sqrt[3]{\sqrt{729 \sin ^4(\theta ) \cos ^2(\theta ) \cos ^2(3 \lambda ) \sin ^2(2 \phi )-108}+27 \sin ^2(\theta ) \cos (\theta ) \cos (3 \lambda ) \sin (2 \phi )}}\nn
			\\
			\omega_0 \to&-\frac{\left(1+i \sqrt{3}\right) \sqrt[3]{\sqrt{729 \sin ^4(\theta ) \cos ^2(\theta ) \cos ^2(3 \lambda ) \sin ^2(2 \phi )-108}+27 \sin ^2(\theta ) \cos (\theta ) \cos (3 \lambda ) \sin (2 \phi )}}{6 \sqrt[3]{2}}
			\\
		&	-\frac{1-i \sqrt{3}}{2^{2/3} \sqrt[3]{\sqrt{729 \sin ^4(\theta ) \cos ^2(\theta ) \cos ^2(3 \lambda ) \sin ^2(2 \phi )-108}+27 \sin ^2(\theta ) \cos (\theta ) \cos (3 \lambda ) \sin (2 \phi )}}\nn
		\end{align}
		and
		\begin{align}
		\psi_l=	\left\{\frac{e^{-i \lambda } \sin (\theta ) \left(\omega_l  \sin (\phi )+e^{3 i \lambda } \cos (\theta ) \cos (\phi )\right)}{\omega_l^2-\sin ^2(\theta ) \cos ^2(\phi )},\frac{e^{-2 i \lambda } \left(\sin ^2(\theta ) \sin (\phi ) \cos (\phi )+e^{3 i \lambda } \omega_l \cos (\theta )\right)}{\omega_l^2-\sin ^2(\theta ) \cos ^2(\phi )},1\right\}
		\end{align}
		where $l=\pm,0$, in the limit of $\lambda \to \pi/6$ after some simplification we get to Eq.\ref{eq.3}.
		
		\section{Density of states}
		\label{DOs.sec}
		Density of states can be calculated by $\rho ({\bf{k}},\omega) =-\frac{1}{\pi}\lim_{\eta \to 0}\text{Im} \text{Tr}  \hat{G}({\bf{k}},i\omega)  $ as follows
		\begin{align}
			\rho({\bf{k}},i\omega=\omega+i\eta)=-\frac{1}{\pi}\lim_{\eta \to 0}\textit{Im} 
			\frac{{3{i\omega ^2} - {{\bf{k}}^2}}}{{i\omega \left( {{i\omega ^2} - {{\bf{k}}^2}} \right)}}
		\end{align}
		
		Total Density of states can be obtained as follows:
		\begin{align}
			\rho(\omega)&=\frac{1}{\frac{4}{3}\pi \Lambda^3}\int_{0}^{\Lambda}{4\pi k^2dk\rho(\bf{k},\omega)} \nonumber
			\\
			&=
			\lim_{\eta \to 0}\textit{Im} \left(- \frac{1}{{\pi i\omega }} + \frac{{6i\omega }}{{\pi {\Lambda ^3}}} + \frac{{3i{\omega ^2}}}{{\pi {\Lambda ^3}}}\left( {\log \left( {\frac{{\Lambda  - i\omega }}{{\Lambda  + i\omega }}} \right) - \log \left( { - 1} \right)} \right)\right)
			\nonumber\\
			&
			=\delta(\omega)+\frac{3\omega^2 \Theta\left(\Lambda-|\omega|\right)}{\Lambda^3}
		\end{align}
		Where $\Lambda$ is the cutoff energy and $\Theta$ is Heaviside function (Fig.\ref{fig:dos}).
		Because we have three bands, the integration of Density of states is 3.
		
		\section{Optical conductivity}
		\label{oc.sec}
		In this section,  we calculate the optical conductivity of the above Hamiltonian with  three fold degeneracy. Our method is based on the Kubo formula. 
		Working in the one-loop approximation, the real part of the
		Kubo formula at finite frequency is
		\footnote{Eq. A1 - Optical conductivity of Weyl semimetals and signatures of the gapped semimetal phase transition}
		\begin{align}
			\sigma_{\alpha\beta}(\Omega)=\frac{e^2\pi}{\hbar\Omega}\int_{-\infty}^{\infty}d\omega[f(\omega)-f(\omega+\Omega)] \int\frac{d^3{\bsk}}{(2\pi)^3}\text{Tr}[\hat v_\alpha \hat A({\bsk},\omega)\hat v_\beta \hat A({\bsk},\omega+\Omega)]
		\end{align}
		where $f(\omega)=[exp(\beta[\omega-\mu])+1]^{-1}$ is the Fermi function. The velocity operator $\hat v_\alpha$ is related to the Hamiltonian via $\hat v_\alpha=\partial \hat H / \partial p_\alpha$. The spectral function $\hat A$ can be obtain by Green function as:
		\begin{align}
			\hat{G}({\bsk},z)=\int_{-\infty}^{\infty}\frac{\hat A({\bsk},\omega)}{z-\omega}d\omega
		\end{align}
		where
		\begin{align}
			\hat{G}^{-1}({\bsk},z)=\hat I z-\hat H
		\end{align}
		where $\hat I$ is identity matrix.  Here we calculate longitudinal conductivity ($\alpha=\beta=x$), and necessary spectral elements of $\hat A$ are $\hat A_{11},\hat A_{12}$  and $\hat A_{22}$
		\begin{align}
			\hat A_{ij}({\bsk},\omega)=-\frac{1}{2\pi}\left(\hat G_{ij}^A({\bsk},i\omega)-\hat G_{ij}^R({\bsk},i\omega)\right)
		\end{align}
		We also set $\hbar=1$ as  ${\bsk}=\hbar {\bf{k}}={\bf{k}}$.
		After some calculation we have

		\begin{align}
			{\hat A_{11}}({\bf{k}},\omega ) 
			&=  \frac{{{k_z}^2}}{{{{\bf{k}}^2}}}\delta (\omega )
			+\frac{1}{2}\left( {1 - \frac{{{k_z}^2}}{{{{\bf{k}}^2}}}} \right)\left( {\delta (\omega  - |{\bf{k}}|) + \delta (\omega  + |{\bf{k}}|)} \right)\\
			{\hat A_{22 }}({\bf{k}},\omega ) 
			&=  \frac{{{k_y}^2}}{{{{\bf{k}}^2}}}\delta (\omega )
			+\frac{1}{2}\left( {1 - \frac{{{k_y}^2}}{{{{\bf{k}}^2}}}} \right)\left( {\delta (\omega  - |{\bf{k}}|) + \delta (\omega  + |{\bf{k}}|)} \right) \\
			{\hat A_{33 }}({\bf{k}},\omega ) 
			&=  \frac{{{k_x}^2}}{{{{\bf{k}}^2}}}\delta (\omega )
			+\frac{1}{2}\left( {1 - \frac{{{k_x}^2}}{{{{\bf{k}}^2}}}} \right)\left( {\delta (\omega  - |{\bf{k}}|) + \delta (\omega  + |{\bf{k}}|)} \right) \\
			{{\hat A}_{12}}({\bf{k}},\omega) 
			&=  - \frac{{{e^{-i\frac{\pi }{3}}}{k_y}{k_z}}}{{{{\bf{k}}^2}}}\delta (\omega ) 
			+ \frac{1}{2}\left( {\frac{{{e^{-i\frac{\pi }{3}}}{k_y}{k_z}}}{{{{\bf{k}}^2}}} - {e^{ i\frac{\pi }{6}}}\frac{{{k_x}}}{{{|{\bf{k}}|}}}} \right)\delta (\omega  + |{\bf{k}}|) 
			+\frac{1}{2}\left( {\frac{{{e^{-i\frac{\pi }{3}}}{k_y}{k_z}}}{{{{\bf{k}}^2}}} + {e^{  i\frac{\pi }{6}}}\frac{{{k_x}}}{{{|{\bf{k}}|}}}} \right)\delta (\omega  - |{\bf{k}}|)
			\\
			{{\hat A}_{13}}({\bf{k}},\omega) 
			&=  - \frac{{{e^{i\frac{\pi }{3}}}{k_x}{k_z}}}{{{{\bf{k}}^2}}}\delta (\omega ) 
			+ \frac{1}{2}\left( {\frac{{{e^{i\frac{\pi }{3}}}{k_x}{k_z}}}{{{{\bf{k}}^2}}} - {e^{ - i\frac{\pi }{6}}}\frac{{{k_y}}}{{{|{\bf{k}}|}}}} \right)\delta (\omega  + |{\bf{k}}|) 
			+ \frac{1}{2}\left( {\frac{{{e^{i\frac{\pi }{3}}}{k_x}{k_z}}}{{{{\bf{k}}^2}}} + {e^{ - i\frac{\pi }{6}}}\frac{{{k_y}}}{{{|{\bf{k}}|}}}} \right)\delta (\omega  - |{\bf{k}}|)
			\\
			{{\hat A}_{23}}({\bf{k}},\omega) 
			&=  - \frac{{{e^{-i\frac{\pi }{3}}}{k_x}{k_y}}}{{{{\bf{k}}^2}}}\delta (\omega ) 
			+ \frac{1}{2}\left( {\frac{{{e^{-i\frac{\pi }{3}}}{k_x}{k_y}}}{{{{\bf{k}}^2}}} - {e^{ i\frac{\pi }{6}}}\frac{{{k_z}}}{{{|{\bf{k}}|}}}} \right)\delta (\omega  + |{\bf{k}}|) 
			+\frac{1}{2}\left( {\frac{{{e^{-i\frac{\pi }{3}}}{k_x}{k_y}}}{{{{\bf{k}}^2}}} + {e^{  i\frac{\pi }{6}}}\frac{{{k_z}}}{{{|{\bf{k}}|}}}} \right)\delta (\omega  - |{\bf{k}}|)
		\end{align}
		and $A_{21}=A_{12}^*$.
		and velocity matrix is
		\begin{align}
			{{\hat v}_\alpha }& = \frac{{\partial \hat H}}{{\partial {p_\alpha }}} 
			\\
			{{\hat v}_x} &= \frac{{\partial \hat H}}{{\partial {p_x}}} = \left( {\begin{array}{*{20}{c}}
					0&a&0\\
					{{a^*}}&0&0\\
					0&0&0
			\end{array}} \right)
		\end{align}
		finally we obtain,
		\begin{align}
			v_x \hat A({\bsk},\omega )= 
			\left( {\begin{array}{*{20}{c}}
					0&a&0\\
					{{a^*}}&0&0\\
					0&0&0
			\end{array}} \right)\left( {\begin{array}{*{20}{c}}
					{{{\hat A}_{11}}}&{{{\hat A}_{12}}}&{{{\hat A}_{13}}}\\
					{{{\hat A}_{21}}}&{{{\hat A}_{22}}}&{{{\hat A}_{23}}}\\
					{{{\hat A}_{31}}}&{{{\hat A}_{32}}}&{{{\hat A}_{33}}}
			\end{array}} \right) = \left( {\begin{array}{*{20}{c}}
					{a{{\hat A}_{21}}}&{a{{\hat A}_{22}}}&{a{{\hat A}_{23}}}\\
					{{a^*}{{\hat A}_{11}}}&{{a^*}{{\hat A}_{12}}}&{{a^*}{{\hat A}_{13}}}\\
					0&0&0
			\end{array}} \right)
		\end{align}
		\begin{align}
			v_x \hat A({\bsk},\omega  + \Omega )=    \left( {\begin{array}{*{20}{c}}
					{a{{\hat A}_{21}}({\bsk},\omega  + \Omega )}&{a{{\hat A}_{22}}({\bsk},\omega  + \Omega )}&{a{{\hat A}_{23}}({\bsk},\omega  + \Omega )}\\
					{{a^*}{{\hat A}_{11}}({\bsk},\omega  + \Omega )}&{{a^*}{{\hat A}_{12}}({\bsk},\omega  + \Omega )}&{{a^*}{{\hat A}_{13}}({\bsk},\omega  + \Omega )}\\
					0&0&0
			\end{array}} \right)
		\end{align}
		\begin{align}
			v_x \hat A({\bsk},\omega ) v_x \hat A({\bsk},\omega  + \Omega )  
			= a{a^*}{{\hat A}_{22}}({\bsk},\omega ){{\hat A}_{11}}({\bsk},\omega  + \Omega ) + {a^*}a{{\hat A}_{11}}({\bsk},\omega ){{\hat A}_{22}}({\bsk},\omega  + \Omega )\nonumber\\
			+ {a^2}{{\hat A}_{21}}({\bsk},\omega ){{\hat A}_{21}}({\bsk},\omega  + \Omega ) + {a^*}^2{{\hat A}_{12}}({\bsk},\omega ){{\hat A}_{12}}({\bsk},\omega  + \Omega )
		\end{align}
		In the next step first of all,  we integrate the first part
		\begin{align}
			&    P_1(\omega,\Omega)=P_1^{D}(\omega,\Omega)+P_1^{\text{IB}}(\omega,\Omega)=\nonumber\\
			&    \int_0^\pi  {\int_0^{2\pi } {\int_0^\Lambda  {{k^2}\sin \theta d\theta d\phi dk} } } \left( {a{a^*}{{\hat A}_{22}}({\bsk},\omega ){{\hat A}_{11}}({\bsk},\omega  + \Omega ) + {a^*}a{{\hat A}_{11}}({\bsk},\omega ){{\hat A}_{22}}({\bsk},\omega  + \Omega )} \right)  
		\end{align}
		Where $P_1^{D}$ is the intraband contribution(Drude) and         
		$P_1^{IB}$ is the  interband contribution of the conductivity.
		\begin{align}
			P_1^{D}(\omega,\Omega)&=
			\frac{{8\pi }}{15}\left( {\frac{{{\Lambda ^3}}}{3}\delta (\omega )\delta (\omega  + \Omega )} \right)+ \frac{{4\pi }}{5}{\omega ^2}
			\left( 
			F_1(\omega,\Omega)
			+ F_2(\omega,\Omega)
			\right) \\
			P_1^{\text{IB}}(\omega,\Omega) &= \frac{{4\pi }}{5}{\omega ^2}
			\left( 
			F_3(\omega,\Omega)
			+ F_4(\omega,\Omega) \right)+ \frac{{16\pi }}{15}\left( {(\omega+\Omega) ^2}\left(
			F_5(\omega,\Omega)
			+F_6(\omega,\Omega)\right)
			+{\omega^2}\left(
			F_7(\omega,\Omega)
			+F_8(\omega,\Omega)\right)
			\right)
		\end{align}
		where 
		\begin{align}
			&    F_1(\omega,\Omega) =     \delta (\Omega )\Theta (\omega )\Theta (\omega  + \Omega )\Theta (\Lambda  - \omega )\Theta (\Lambda  - (\omega  + \Omega ))\nn\\
			&F_2(\omega,\Omega) = \delta (\Omega )\Theta ( - \omega )\Theta ( - \omega  - \Omega )\Theta (\Lambda  + \omega )\Theta (\Lambda  + (\omega  + \Omega )) \nn\\
			&F_3(\omega,\Omega) =\delta (2\omega  + \Omega )\Theta (\omega )\Theta ( - \omega  - \Omega )\Theta (\Lambda  - \omega )\Theta (\Lambda  + (\omega  + \Omega )) \nn\\
			&F_4(\omega,\Omega) = \delta (2\omega  + \Omega )\Theta ( - \omega )\Theta (\omega  + \Omega )\Theta (\Lambda  + \omega )\Theta (\Lambda  - (\omega  + \Omega )) \nn\\
			&F_5(\omega,\Omega)= \delta (\omega )\Theta (\omega  + \Omega )\Theta (\Lambda  - (\omega  + \Omega ))\nn\\
			&F_6(\omega,\Omega) =  \delta (\omega )\Theta ( - \omega  - \Omega )\Theta (\Lambda  + (\omega  + \Omega )) \nn\\
			&F_7(\omega,\Omega)= \delta (\omega  + \Omega )\Theta (\omega )\Theta (\Lambda  - \omega ) \nn\\
			&F_8(\omega,\Omega)= \delta (\omega  + \Omega )\Theta ( - \omega )\Theta (\Lambda  + \omega )
		\end{align}
		The integrate of second part is
		\begin{align}
			&    P_2(\omega,\Omega)=P_2^{D}(\omega,\Omega)+P_2^{IB}(\omega,\Omega)=\nonumber\\
			&    \int_0^\pi  {\int_0^{2\pi } {\int_0^\Lambda  {{k^2}\sin \theta d\theta d\phi dk} } } \left( {{a^2}{{\hat A}_{21}}({\bsk},\omega ){{\hat A}_{21}}({\bsk},\omega  + \Omega ) + {a^*}^2{{\hat A}_{12}}({\bsk},\omega ){{\hat A}_{12}}({\bsk},\omega  + \Omega )} \right) 
		\end{align}
		and
		\begin{align}
			P_2^{D}(\omega,\Omega)=&-
			\frac{{ 8\pi }}{{15}}\left( {\frac{{{\Lambda ^3}}}{3}\delta (\omega )\delta (\omega  + \Omega )} \right)
			+ \frac{{8\pi }}{{15}}{\omega ^2}\left( {\begin{array}{*{20}{l}}
					F_1(\omega+\Omega)+    F_2(\omega+\Omega)
			\end{array}} \right)\nonumber\\
			\\
			P_2^{\text{IB}}(\omega,\Omega)=&-
			\frac{{  4\pi }}{5}{\omega ^2}\left( {\begin{array}{*{20}{l}}
					F_3(\omega+\Omega)+     F_4(\omega+\Omega)    
			\end{array}} \right)\nonumber\\&
			+ \frac{{4\pi }}{{15}}
			\left( {(\omega+\Omega) ^2}\left(
			F_5(\omega,\Omega)
			+F_6(\omega,\Omega)\right)
			+{\omega^2}\left(
			F_7(\omega,\Omega)
			+F_8(\omega,\Omega)\right)
			\right)
		\end{align}
		In the last step, we integrate part 1 and part 2 of intraband part,
		\begin{align}
			\sigma_{xx}^D(\Omega)&=\frac{\sigma_0\pi}{\Omega}\frac{1}{(2\pi)^3}\int_{-\infty}^{\infty}d\omega(f(\omega)-f(\omega+\Omega))(P_1^D(\omega,\Omega)+P_2^D(\omega,\Omega)) \nonumber \\
			&    
			=\frac{4\pi}{3}\frac{\sigma_0\pi}{\Omega}\frac{1}{(2\pi)^3}\int_{ - \infty }^\infty  {d\omega \left( {\Omega \frac{{ - \partial f(\omega )}}{{\partial \omega }}} \right){\omega ^2}} \left(     F_1(\omega+\Omega)+    F_2(\omega+\Omega) \right) \\
			&    = \frac{4\pi}{3}\frac{\sigma_0\pi}{\Omega}\frac{1}{(2\pi)^3}\delta(\Omega)\left. {\frac{{\beta \omega \left( {\frac{{{{\rm{e}}^{\beta (\omega+\mu) }}\beta \omega }}{{1 + {{\rm{e}}^{\beta (\omega+\mu) }}}} - 2\log \left( {1 + {{\rm{e}}^{\beta (\omega+\mu) }}} \right)} \right) - 2\text {Li} _ 2\left (-e^{\beta(\omega+\mu)} \right)}}{{{\beta ^2}}}} \right|_{-\Lambda}^{\Lambda}
		\end{align}
		where $\text {Li} _ n(z)$ is Polylogarithm  function. for $|z|\gg 1$ (for $\beta(\Lambda+\mu) \to \infty$,),
		\begin{align}
			\lim_{z \to \infty}{\text {Li} _ 2\left (-e^{z} \right)}=- \frac{{{{z }^2}}}{2} - \frac{{{\pi ^2}}}{6} + \mathcal{O} \left( {\frac{1}{{{{z }^2}}}} \right)
		\end{align} 
		As a result for $\lim_{\Lambda \to \infty} {\sigma_{xx}^D(\Omega)}$ we have
		\begin{align}
			\sigma_{\alpha\beta}^D(\Omega)&=\sigma_0\pi\frac{4\pi}{3}\frac{1}{(2\pi)^3}\left(\mu^2+\frac{\pi^2}{3}k_B^2T^2\right)\delta(\Omega) 
			\\
			&=
			\frac{ \sigma_0}{6\pi}\left(\mu^2+\frac{\pi^2}{3}k_B^2T^2\right)\delta(\Omega). 
		\end{align}
		The total optical spectral weight under the Drude (defined
		as
		$W_D=\int_{0^+}^{\infty}\sigma_{\alpha\beta}^D(\Omega)d\Omega$)
		is
		\begin{align}
			W_D=\frac{ \sigma_0}{6\pi}\left(\mu^2+\frac{\pi^2}{3}k_B^2T^2\right)
		\end{align}
		We integrate part 1 and part 2 of interband part, for $\Lambda \to \infty$,  we obtain,
		\begin{align}
			\sigma_{xx}^{\text{IB}}(\Omega)&=\frac{\sigma_0\pi}{\Omega}\frac{1}{(2\pi)^3}\int_{-\infty}^{\infty}d\omega\left(f(\omega)-f(\omega+\Omega)\right)\left(P_1^{\text{IB}}(\omega,\Omega)+P_2^{\text{IB}}(\omega,\Omega)\right) \nonumber \\
			&    
			=\frac{{4\pi }}{3}\frac{\sigma_0\pi}{\Omega}\frac{1}{(2\pi)^3}\int_{ - \infty }^\infty  {d\omega \left(f(\omega)-f(\omega+\Omega)\right)  } \left( {(\omega+\Omega) ^2}\left(
			F_5(\omega,\Omega)
			+F_6(\omega,\Omega)\right)
			+{\omega^2}\left(
			F_7(\omega,\Omega)
			+F_8(\omega,\Omega)\right)
			\right)
		\end{align} 
		After some calculation,
		\begin{align}
			\sigma_{xx}^{\text{IB}}(\Omega)    = \frac{{\sigma_0}}{6\pi}\Omega\left(f(-\Omega)-f(\Omega)\right).
		\end{align}
		\end{widetext}
	
	\bibliographystyle{apsrev4-1}
	\bibliography{Refs}
\end{document}